\documentclass[aps,pra,twocolumn,superscriptaddress,showpacs,nofootinbibfloatfix,amsmath,amsfonts,amssymb]{revtex4-2}%
\usepackage{amsmath,amsfonts,amssymb,color}
\usepackage{amsthm}
\usepackage{leftidx}
\usepackage{graphicx}
\usepackage{xcolor}
\usepackage{dcolumn}
\usepackage{bm}
\usepackage{epstopdf}
\usepackage{epsfig}
\usepackage{environ}
\usepackage{pdfcomment}

\usepackage{multirow}
\usepackage{setspace}
\usepackage{color}

\usepackage{float}
\usepackage[T1]{fontenc}
\usepackage[latin9]{inputenc}
\usepackage{setspace}
\usepackage{esint}

\usepackage{wasysym}

\begin{document}


\title{Non-Hermitian topological Euler insulators}

\author{Longwen Zhou}
\email{zhoulw13@u.nus.edu}
\affiliation{%
	College of Physics and Optoelectronic Engineering, Ocean University of China, Qingdao 266100, China 
}
\affiliation{%
	Qingdao Key Laboratory of Advanced Optoelectronics, Qingdao 266100, China
}
\affiliation{%
	Engineering Research Center of Advanced Marine Physical Instruments and Equipment of MOE, Qingdao 266100, China
}

\date{\today}

\begin{abstract}
Topological Euler insulators emerge in multiband systems with real
Bloch Hamiltonians and wavefunctions. Their fragile topologies are
characterized by the Euler class of degenerate bands and protected
by the $PT$ or $C_{2}T$ symmetry in two dimensions, which go beyond
the tenfold $K$-theory classification of topological matter. In this
work, we extend the conception of topological Euler insulators to
non-Hermitian systems and propose a theoretical framework to unlock
their nontrivial Euler topology. Focusing on two-dimensional, three-band
non-Hermitian lattice models with symmetric Hamiltonians, we formulate
a comprehensive description of their topological Euler bands, entanglement
spectrum and bulk-boundary correspondence. Three typical models of
non-Hermitian Euler insulators are constructed and investigated explicitly
to illustrate our theory. Unique topological phase transitions and
anomalous edge-band overlaps with non-Hermitian origins are further
identified. Our study establishes the presence of topological Euler
bands in non-Hermitian systems and unveils their intriguing physical
characteristics, thereby broadening the existing territory of topological
matter in non-Hermitian open systems.
\end{abstract}

\pacs{}
\keywords{}
\maketitle

\section{Introduction\label{sec:Int}}

Topological Euler insulators have recently emerged as a captivating
paradigm in condensed matter physics, describing multiband systems
with symmetric Bloch Hamiltonians and real wavefunctions. Unlike traditional
topological phases characterized by $K$-theory \cite{Tenfold}, the topology of Euler
insulators is classified by the non-Abelian Euler class $e_{12}$,
an integer invariant defined over degenerate pairs of Bloch bands
\cite{TEI01,TEI02,TEI03,TEI04,TEI05,TEI06,TEI07,TEI08,TEI09,TEI10}.
Protected by space-time inversion symmetries such as $PT$ or $C_{2}T$,
these fragile topological phases exhibit rich multigap phenomena,
non-Abelian braiding dynamics, and unique bulk-boundary correspondence
\cite{TEI11,TEI12,TEI13,TEI14,TEI15,TEI16,TEI17,TEI18,TEI19,TEI20,TEI21,TEI22,TEI23,TEI24,TEI25,TEI26,TEI27,TEI28,TEI29,TEI30,TEI31,TEI32,TEI33,TEI34,TEI35,TEI36,TEI37,TEI38,TEI39,TEI40,TEI41,TEI42,TEI43,TEI44,TEI45,TEI46}.
While extensively studied in Hermitian settings such as trapped ions
\cite{TEI10}, ultracold atoms \cite{TEI06,TEI17}, acoustic metamaterials
\cite{TEI22} and electronic structures \cite{TEI02,TEI05}, the fate
and character of Euler topology in non-Hermitian systems remain largely
unexplored. Notably, as gain, loss, and nonreciprocal couplings could
break key Hermiticity constraints, the very existence of topological
Euler insulators in non-Hermitian settings constitutes a challenging
issue to address.

In the meantime, non-Hermitian physics has fundamentally reshaped
our understanding of open quantum systems, giving rise to unique phenomena
such as $PT$-symmetry breaking \cite{PTSym01,PTSym02,PTSym03}, point/line
gap topology \cite{NHTP01,NHTP02,NHTP03}, exceptional points \cite{EP01,EP02,EP03},
and non-Hermitian skin effects \cite{NHSE01,NHSE02,NHSE03}. However,
extending topological Euler invariants to non-Hermitian settings presents
a fundamental theoretical challenge, i.e., non-Hermiticity generally
forces wavefunctions into complex spaces and destroys the realness
conditions that protect standard Euler classes. Unlocking Euler topology
in non-Hermitian systems requires establishing a generalized formulation
for topological invariants, clarifying how non-Hermiticity reshapes
phase transitions, and determining whether robust bulk-boundary correspondence
and edge-state signatures survive. Addressing these questions is crucial
to expanding the boundaries of multigap topological matter into open
nonequilibrium systems, hereby motivating the present study.

In this work, we establish an overall theoretical framework for non-Hermitian
topological Euler insulators. We focus on two-dimensional (2D), three-band
non-Hermitian lattice models with transposition-invariant Hamiltonians,
and develop a systematic characterization of their topological Euler
bands, entanglement spectrum, and bulk-boundary correspondence. Our
construction starts from a generating vector ${\bf h}({\bf k})$ whose
components could be complex, from which one obtains a three-band Bloch
Hamiltonian $H({\bf k})$ with a symmetric matrix structure. To reveal
its topological phases, we introduce a generalized Euler class defined
in biorthonormal basis, demonstrate its quantization in gapped non-Hermitian
phases, and establish its connection to the winding number of Wilson
loop spectra and the Chern number of an associated two-band model.

To illustrate our theory explicitly, we construct and investigate
three representative models of non-Hermitian Euler insulators. The
first originates from the Qi-Wu-Zhang Chern insulator with complex
chemical potential, uncovering non-Hermiticity-induced phase transitions
mediated by gapless intermediate phases. The second incorporates next-nearest-neighbor
hoppings in its generating vector, demonstrating that non-Hermitian
effects can sustain topological Euler insulators with large Euler
class $e_{12}=4$. The third model, rooted in a square-lattice realization
of the Haldane Chern insulator, unveils anomalous edge-band
overlaps in both energy and entanglement spectra with non-Hermitian
origins. Across all three models, we establish a bulk-boundary correspondence
in which the absolute value of the Euler class counts the number of
quadratic touchings (or their non-Hermitian generalizations) between
edge bands.

The remainder of this paper is organized as follows. In Sec.~\ref{sec:Mod},
we introduce the general construction of our non-Hermitian lattice
model and discuss its basic properties. In Sec.~\ref{sec:The}, we
propose the theoretical framework for characterizing non-Hermitian
Euler topology, including the spectral gap function, generalized Euler
class, Wilson loop winding numbers, entanglement spectrum, and bulk-boundary
correspondence. In Sec.~\ref{sec:Res}, we apply this framework to
three concrete models, presenting detailed numerical results on their
phase diagrams, topological invariants, entanglement signatures, and
edge-state spectra. We conclude with a discussion of our findings
and an outlook for future directions in Sec.~\ref{sec:Sum}.

\section{Model\label{sec:Mod}}

We first introduce a class of 2D non-Hermitian lattice model, whose
gapped phases could be characterized by the Euler class. Our construction
starts with a three-component generating vector ${\bf h}({\bf k})=[h_{x}({\bf k}),h_{y}({\bf k}),h_{z}({\bf k})]$,
where ${\bf k}=(k_{x},k_{y})$. $k_{x}\in[-\pi,\pi)$ and $k_{x}\in[-\pi,\pi)$
represent the quasimomenta along $x$ and $y$ directions of the lattice.
All components of ${\bf h}({\bf k})$ are $2\pi$ periodic in both
$k_{x}$ and $k_{y}$. To make the resulting system non-Hermitian,
we choose at least one component of ${\bf h}({\bf k})$ to be complex-valued
at some ${\bf k}$. Besides this and the momentum-space periodicity,
we do not assume further symmetry constraints on ${\bf h}({\bf k})$.

Using the generating vector ${\bf h}({\bf k})$, we could obtain a
three-band Bloch Hamiltonian $H({\bf k})$, whose generic form reads
\begin{equation}
	H({\bf k})=[{\bf h}({\bf k})\cdot{\bf h}({\bf k})]|u_{3}({\bf k})\rangle\langle\tilde{u}_{3}({\bf k})|.\label{eq:Hk}
\end{equation}
Here, the normalized right eigenvector $|u_{3}({\bf k})\rangle$ of
$H({\bf k})$ is given by
\begin{equation}
	|u_{3}({\bf k})\rangle=\frac{1}{\sqrt{{\bf h}({\bf k})\cdot{\bf h}({\bf k})}}\begin{bmatrix}h_{x}({\bf k})\\
		h_{y}({\bf k})\\
		h_{z}({\bf k})
	\end{bmatrix},\label{eq:u3k}
\end{equation}
and the associated left eigenvector $\langle\tilde{u}_{3}({\bf k})|\equiv[|u_{3}({\bf k})\rangle]^{\top}$,
where $\top$ denotes transpose. It can be verified that $H({\bf k})|u_{3}({\bf k})\rangle=E_{3}({\bf k})|u_{3}({\bf k})\rangle$
and $\langle\tilde{u}_{3}({\bf k})|H({\bf k})=\langle\tilde{u}_{3}({\bf k})|E_{3}({\bf k})$,
where the dispersion of band $3$ reads $E_{3}({\bf k})={\bf h}({\bf k})\cdot{\bf h}({\bf k})$.
$H({\bf k})$ has two other flat bands $E_{1}({\bf k})$ and $E_{2}({\bf k})$
at zero energy, i.e., $E_{1}({\bf k})=E_{2}({\bf k})=0$ for all ${\bf k}$
in the first Brillouin zone (BZ). Their right and left eigenvectors,
which are energetically degenerate, can be constructed as
\begin{alignat}{1}
	|u_{1}({\bf k})\rangle &= \frac{1}{\sqrt{h_{x}^{2}+h_{y}^{2}}}\begin{bmatrix}h_{y}\\
		-h_{x}\\
		0
	\end{bmatrix},\label{eq:u1k}\\
	|u_{2}({\bf k})\rangle &= \frac{1}{\sqrt{E_{3}({\bf k})}\sqrt{h_{x}^{2}+h_{y}^{2}}}\begin{bmatrix}h_{x}h_{z}\\
		h_{y}h_{z}\\
		-h_{x}^{2}-h_{y}^{2}
	\end{bmatrix},\label{eq:u2k}
\end{alignat}
$\langle\tilde{u}_{1}({\bf k})|=[|u_{1}({\bf k})\rangle]^{\top}$
and $\langle\tilde{u}_{2}({\bf k})|=[|u_{2}({\bf k})\rangle]^{\top}$.
These eigenvectors satisfy the orthogonality and completeness relations
of biorthonormal basis \cite{Brody2014}, i.e.,
\begin{equation}
	\langle\tilde{u}_{\alpha}({\bf k})|u_{\beta}({\bf k})\rangle=\delta_{\alpha\beta},\quad\sum_{\alpha=1}^{3}|u_{\alpha}({\bf k})\rangle\langle\tilde{u}_{\alpha}({\bf k})|=\mathbb{I}_{3},\label{eq:bio}
\end{equation}
where $\alpha,\beta=1,2,3$, and $\mathbb{I}_{3}$ denotes the $3\times3$
identity. The sets of energy bands $\{E_{\alpha}({\bf k})|\alpha=1,2,3\}$
and eigenvectors $\{|u_{\alpha}({\bf k})\rangle,|\tilde{u}_{\alpha}({\bf k})\rangle|\alpha=1,2,3\}$
yield a complete description of all bulk phases in the system described
by $H({\bf k})$.

The Hamiltonian $H({\bf k})$ in Eq.~(\ref{eq:Hk}) has the explicit matrix form
\begin{equation}
	H({\bf k})=\begin{bmatrix}h_{x}^{2} & h_{x}h_{y} & h_{x}h_{z}\\
		h_{x}h_{y} & h_{y}^{2} & h_{y}h_{z}\\
		h_{x}h_{z} & h_{y}h_{z} & h_{z}^{2}
	\end{bmatrix}.\label{eq:HkM}
\end{equation}
This Hamiltonian matrix is transposition-symmetric, i.e.,
$H({\bf k})=H^{\top}({\bf k})$. In the Hermitian limit (i.e., $h_{x}$,
$h_{y}$ and $h_{z}$ are all real functions of ${\bf k}$), this
symmetry further implies that $H({\bf k})$ is a real Hermitian
matrix, whose eigenstates are all real vectors. In our non-Hermitian
case (i.e., $h_{x}$, $h_{y}$ and $h_{z}$ can be complex), the eigenvectors of $H({\bf k})$ could also become
complex vectors, violating the realness conditions required for the
definition of standard Euler class. However, we will see that the
transposition-symmetric structure of $H({\bf k})$ still allow its
biorthonormal eigenvectors to carry quantized Euler class in topologically
nontrivial non-Hermitian gapped phases. Therefore, the realness of
the eigenvectors of $H({\bf k})$ is not a necessary condition for
the system to show nontrivial Euler topology in non-Hermitian cases.

Before introducing the theoretical framework to characterize non-Hermitian
Euler topology, let us elaborate a bit more on the motivation behind
our model construction. The generating vector ${\bf h}({\bf k})$
can also be used to define a lattice model $h({\bf k})={\bf h}({\bf k})\cdot\boldsymbol{\sigma}$
with two energy bands $\varepsilon_{\pm}({\bf k})=\pm\varepsilon({\bf k})=\pm\sqrt{{\bf h}\cdot{\bf h}}$,
where $\boldsymbol{\sigma}=(\sigma_{x},\sigma_{y},\sigma_{z})$ contains
three Pauli matrices as its components. The right and left eigenvectors
for the band $\varepsilon_{-}({\bf k})$ of Hamiltonian $h({\bf k})$
are given by
\begin{alignat}{1}
	|\psi_{-}^{R}\rangle &= \frac{1}{\sqrt{2\varepsilon(\varepsilon+h_{z})}}\begin{bmatrix}h_{x}-ih_{y}\\
		-\varepsilon-h_{z}
	\end{bmatrix},\label{eq:psiR1}\\
	\langle\psi_{-}^{L}| &= \frac{1}{\sqrt{2\varepsilon(\varepsilon+h_{z})}}\begin{bmatrix}h_{x}+ih_{y} & -\varepsilon-h_{z}\end{bmatrix}.\label{eq:psiL1}
\end{alignat}
In biorthonormal basis, the Berry curvature of the band $\varepsilon_{-}({\bf k})$
reads
\begin{equation}
	F_{xy}=i\left[\langle\partial_{x}\psi_{-}^{L}|\partial_{y}\psi_{-}^{R}\rangle-\langle\partial_{y}\psi_{-}^{L}|\partial_{x}\psi_{-}^{R}\rangle\right].\label{eq:BC}
\end{equation}
A straightforward yet tedious calculation yields
\begin{equation}
	F_{xy}=\frac{1}{2}\frac{{\bf h}\cdot(\partial_{k_{x}}{\bf h}\times\partial_{k_{y}}{\bf h})}{({\bf h}\cdot{\bf h})^{3/2}}.\label{eq:BC2}
\end{equation}
Therefore, the Chern number $\mathfrak{c}_{-}$ of energy band $\varepsilon_{-}({\bf k})$
turns out to be
\begin{equation}
	\mathfrak{c}_{-}=\frac{1}{2\pi}\int_{{\rm BZ}}{\rm d}^{2}{\bf k}F_{xy}=\frac{1}{4\pi}\int_{{\rm BZ}}{\rm d}^{2}{\bf k}\frac{{\bf h}\cdot(\partial_{k_{x}}{\bf h}\times\partial_{k_{y}}{\bf h})}{({\bf h}\cdot{\bf h})^{3/2}},\label{eq:CN}
\end{equation}
which has been shown to possess integer values for separable non-Hermitian
Chern bands \cite{NHCI}. In the next section, we will see that
there exists a direct connection between the Chern number $\mathfrak{c}_{-}$
and the Euler class of the three-band model $H({\bf k})$, thereby
ensuring its gapped topological phases to carry nontrivial integer
Euler class.

\section{Theory\label{sec:The}}

In this section, we propose a theoretical framework to describe the
bulk spectra, topological phases, entanglement properties and edge
states of non-Hermitian Euler insulators. Although the theory is formulated
with the three-band Hamiltonian $H({\bf k})$ in mind, the obtained
expressions are principally generalizable to multiple band systems.

First, to identify a phase transition accompanied by the gap closing
in the bulk spectrum of $H({\bf k})$, we introduce the spectral gap
function $\Delta$, which is defined as 
\begin{equation}
	\Delta=\min_{{\bf k}\in{\rm BZ}}|E_{3}({\bf k})|.\label{eq:GF}
\end{equation}
Here, since the other two degenerate bands $E_{1}$ and $E_{2}$ of
$H({\bf k})$ are pinned at zero energy, the gap function $\Delta$
measures the minimal distance between the band $E_{3}({\bf k})$ and
the origin on the complex plane. Therefore, when $\Delta>0$, the
bands $E_{1,2}$ and $E_{3}({\bf k})$ are separated by a spectrum
gap. We refer to this case as a non-Hermitian insulating phase (assuming
that the bands $E_{1,2}$ are uniformly filled). When $\Delta=0$,
the bands $E_{3}({\bf k})$ and $E_{1,2}$ are either touched or overlapped,
yielding a critical point of phase transition or a gapless phase.
The loci of $\Delta$ where it changes from positive to zero in the
parameter space of $H$ then determine the phase boundaries of the
system. In this work, we will mainly focus on the topological characterization
of non-Hermitian gapped phases with $\Delta>0$.

Next, to characterize the nontrivial topology of the system, we introduce
a generalized form of Euler class $e_{12}$ for the two degenerate
bands $E_{1}$ and $E_{2}$, which is given by
\begin{equation}
	e_{12}=\frac{1}{2\pi}\int_{{\rm BZ}}{\rm d}^{2}{\bf k}A_{12}({\bf k}),\label{eq:EC}
\end{equation}
where $A_{12}=\langle\partial_{k_{x}}\tilde{u}_{1}({\bf k})|\partial_{k_{y}}u_{2}({\bf k})\rangle-\langle\partial_{k_{y}}\tilde{u}_{1}({\bf k})|\partial_{k_{x}}u_{2}({\bf k})\rangle$.
Compared to the standard definition of Euler class for Hermitian bands,
the difference here is the incorporation of biorthonormal basis, which
encodes the key distinction between Hermitian and non-Hermitian descriptions.
Inserting Eqs.~(\ref{eq:u1k}), (\ref{eq:u2k}), and their corresponding
left eigenvectors into Eq.~(\ref{eq:EC}), we obtain after some algebraic
manipulations that 
\begin{equation}
	e_{12}=\frac{1}{2\pi}\int_{{\rm BZ}}{\rm d}^{2}{\bf k}\frac{{\bf h}\cdot(\partial_{k_{x}}{\bf h}\times\partial_{k_{y}}{\bf h})}{({\bf h}\cdot{\bf h})^{3/2}}.\label{eq:PI}
\end{equation}
In this form, the $e_{12}$ corresponds to an extension of the Pontryagin
number \cite{TEI10}, i.e., twice of the winding number of ${\bf n}({\bf k})={\bf h}/\sqrt{{\bf h}\cdot{\bf h}}$
over the BZ, with the ${\bf n}({\bf k})$ being allowed to be a complex
vector here. Importantly, we notice from Eq.~(\ref{eq:CN}) that 
\begin{equation}
	e_{12}=2\mathfrak{c}_{-}.\label{eq:e12CN}
\end{equation}
Therefore, for a gapped topological phase of $H({\bf k})$, the quantization
of Euler class $e_{12}$ is guaranteed by the integer quantization
of $\mathfrak{c}_{-}$ for a separable Chern band. The Euler class
$e_{12}$ thus forms a well-defined, even-integer topological index
that will be used to characterize Euler insulator phases in our non-Hermitian
system.

In the study of Euler insulators, the winding number of Wilson loop
spectrum can also be used to characterize Euler topology
\cite{TEI10}. Within the biorthogonal formalism, the Wilson loop
is generalizable to non-Hermitian settings. For example,
the Wilson loop $W_{y}$ along the $y$ direction of the lattice has
the matrix element 
\begin{equation}
	[W_{y}]_{\alpha\beta}=\langle\tilde{u}_{\alpha}({\bf k}_{0})|\prod_{{\bf k}_{\ell}}P_{{\rm occ}}({\bf k}_{\ell})|u_{\beta}({\bf k}_{0})\rangle.\label{eq:Wy}
\end{equation}
Here, ${\bf k}_{0}=(k_{x},k_{y})$ represents the base point of the
Wilson loop. ${\bf k}_{\ell}=(k_{x},k_{y}+2\pi\ell/N_{y})$, where
$\ell=1,...,N_{y}$, and $N_{y}$ denotes the total number of unit
cells along the $y$ direction. $P_{{\rm occ}}({\bf k}_{\ell})$ represents
the projector onto the occupied bands of the system at each ${\bf k}_{\ell}$.
For our three-band Hamiltonian $H({\bf k})$, we have $\alpha,\beta=1,2,3$,
and the projector $P_{{\rm occ}}({\bf k}_{\ell})$ is explicitly given
by
\begin{equation}
	P_{{\rm occ}}({\bf k}_{\ell})=\mathbb{I}_{3}-|u_{3}({\bf k}_{\ell})\rangle\langle\tilde{u}_{3}({\bf k}_{\ell})|.\label{eq:Pocc}
\end{equation}
Compared to the definition of Wilson loop in Hermitian systems, we
notice that the only difference here is the application of biorthonormal
basis. The topological winding number of Wilson loop can then be obtained
from the spectrum of the path projector
\begin{equation}
	P(k_{x})=\prod_{{\bf k}_{\ell}}P_{{\rm occ}}({\bf k}_{\ell}).\label{eq:P}
\end{equation}
Since the state $|u_{3}({\bf k}_{\ell})\rangle$ is excluded from
$P_{{\rm occ}}({\bf k}_{\ell})$ at each ${\bf k}_{\ell}$, one of
the eigenvalue of $P(k_{x})$ should be zero at every $k_{x}\in[-\pi,\pi)$.
The phase factors $\theta_{y}^{\pm}(k_{x})$ of the other two eigenvalues
of $P(k_{x})$ form a pair of phase bands vs $k_{x}$. Due to the
periodicity of $P(k_{x})$ in momentum space, the phase bands $\theta_{y}^{\pm}(k_{x})$
must carry integer-quantized winding numbers over the BZ whenever
the energy spectrum gap $\Delta>0$. Formally, we can express these
winding numbers as
\begin{equation}
	w_{y}^{\pm}=\frac{1}{2\pi}\int_{-\pi}^{\pi}{\rm d}k_{x}\partial_{k_{x}}\theta_{y}^{\pm}(k_{x}),\label{eq:wypm}
\end{equation}
where $w_{y}^{+}=-w_{y}^{-}$. The Wilson loop $W_{x}$ along the
$x$ direction and its associated winding number can be obtained following
the same routine. In Hermitian cases, the Wilson loop winding number
is equivalent to the Euler class \cite{TEI07}, in the sense that
$|e_{12}|=|w_{y}^{\pm}|$. In our case, we will see from explicit
examples that this equivalence holds also for non-Hermitian gapped
phases with nontrivial Euler class topology.

The entanglement spectrum (ES) reflects the quantum correlation among
different subsystems of a given many-body system across its selected
entanglement cuts. It contains important information about the topology,
criticality and bulk-boundary correspondence of topological phases.
For a system made up of free fermions, the ES can be extracted from
the spectrum of single-particle correlation matrix \cite{ES1}. Let
us first decompose our 2D system into two subsystems A and B. The
system is prepared in an insulating ground state $|\Psi_{0}\rangle$,
in which its two degenerate bulk bands at zero energy are uniformly
filled. The reduced density matrix $\rho_{{\rm A}}$ of subsystem
A can then be obtained by tracing over all degrees of freedom belonging
to the subsystem B from the biorthonormal ground state projector $|\Psi_{0}\rangle\langle\tilde{\Psi}_{0}|$,
i.e.,
\begin{equation}
	\rho_{{\rm A}}={\rm Tr}_{{\rm B}}\left(|\Psi_{0}\rangle\langle\tilde{\Psi}_{0}|\right)=\frac{1}{Z_{{\rm A}}}e^{-H_{{\rm A}}},\label{eq:RDM}
\end{equation}
where $Z_{{\rm A}}\equiv{\rm Tr}\left(e^{-H_{{\rm A}}}\right)$ corresponds
to a normalization factor. The term $H_{{\rm A}}$ on the exponential
is usually called the entanglement Hamiltonian, whose spectrum $\{\zeta_{j}\}$
gives the ES. An equivalent expression of the ES can be obtained from
$\{\zeta_{j}\}$ through a one-to-one mapping, yielding
\begin{equation}
	\{\xi_{j}\}=\left\{ \frac{1}{e^{\zeta_{j}}+1}\right\} .\label{eq:ES}
\end{equation}
In this definition, the range of ES is confined to $[0,1]$ for Hermitian
systems. The set $\{\xi_{j}\}$ is further equivalent to the eigenspectrum
of single-particle correlator restricted to the subsystem A for Gaussian
states of free fermions \cite{ES1}. In non-Hermitian cases, the ES
$\{\xi_{j}\}$ may take values beyond the range $[0,1]$ since the
spectrum $\{\zeta_{j}\}$ could be complex. Nevertheless, we can still
deduce the ES by diagonalizing the single-particle correlation matrix
expressed in biorthonormal basis (see Refs.~\cite{ES2,ES3,ES4} for
derivation details). Here we outline the essential steps leading to
this equivalence.

Let ${\bf r},{\bf r}'$ and $\sigma,\sigma'$ be the indices of unit
cells and internal degrees of freedom, respectively, with ${\bf r},{\bf r}'$
belonging to the subsystem A. The single-particle correlator $C_{{\rm A}}$
within subsystem A has the matrix element
\begin{equation}
	[C_{{\rm A}}]_{{\bf r}\sigma,{\bf r}'\sigma'}={\rm Tr}\left(\rho_{{\rm A}}c_{{\bf r}'\sigma'}^{\dagger}c_{{\bf r}\sigma}\right)=\langle\tilde{\Psi}_{0}|c_{{\bf r}'\sigma'}^{\dagger}c_{{\bf r}\sigma}|\Psi_{0}\rangle,\label{eq:CM1}
\end{equation}
where ${\bf r},{\bf r}'\in{\rm A}$, and $c_{{\bf r}'\sigma'}^{\dagger}$ ($c_{{\bf r}\sigma}$) creates
(annihilates) a fermion with internal index $\sigma'$ ($\sigma$)
at the unit cell ${\bf r}'$ (${\bf r}$). In the second equality,
we notice that the $C_{{\rm A}}$ can be obtained by restricting the
degrees of freedom ${\bf r},{\bf r}'$ and $\sigma,\sigma'$ to subsystem
A. With this in mind, we can first compute the correlation matrix
$C$ of the full system ${\rm A+B}$. Under periodic boundary conditions
(PBC) along both the $x$ and $y$ directions, we can expand each
term on the right hand side of Eq.~(\ref{eq:CM1}) in momentum space
as $c_{{\bf r}\sigma}=\frac{1}{\sqrt{N}}\sum_{{\bf k}\in{\rm BZ}}e^{i{\bf k}\cdot{\bf r}}c_{{\bf k}\sigma}$,
$c_{{\bf r}'\sigma'}^{\dagger}=\frac{1}{\sqrt{N}}\sum_{{\bf k}'\in{\rm BZ}}e^{-i{\bf k}'\cdot{\bf r}'}c_{{\bf k}'\sigma'}^{\dagger}$,
\begin{alignat}{1}
	|\Psi_{0}\rangle= & \prod_{{\bf k}\in{\rm BZ},\alpha=1,2}\left[\sum_{\sigma}u_{\alpha,\sigma}({\bf k})c_{{\bf k}\sigma}^{\dagger}\right]|\emptyset\rangle,\label{eq:GSR}\\
	\langle\tilde{\Psi}_{0}|= & \prod_{{\bf k}\in{\rm BZ},\alpha=1,2}\langle\emptyset|\left[\sum_{\sigma}u_{\alpha,\sigma}({\bf k})c_{{\bf k}\sigma}\right].\label{eq:GSL}
\end{alignat}
Here, $N=N_{x}N_{y}$ counts the total number of unit cells in the
full system ${\rm A+B}$. ${\bf r}=(m,n)$ represents the unit cell
index. $\alpha=1,2$ include the indices of occupied bands. $|\emptyset\rangle$
denotes the vacuum state. $u_{\alpha,\sigma}({\bf k})$ represents
the $\sigma$th ($\sigma=1,2,3$) entry of the right and left eigenvectors
$|u_{\alpha}({\bf k})\rangle$ and $\langle\tilde{u}_{\alpha}({\bf k})|$
of $H({\bf k})$. Plugging all these expansions into the second equality
of Eq.~(\ref{eq:CM1}), we arrive at the matrix element of $C$, i.e.,
\begin{equation}
	[C]_{{\bf r}\sigma,{\bf r}'\sigma'}=\frac{1}{N}\sum_{\alpha=1,2}\sum_{{\bf k},{\bf k}'\in{\rm BZ}}e^{i({\bf k}\cdot{\bf r}-{\bf k}'\cdot{\bf r}')}u_{\alpha,\sigma'}({\bf k}')u_{\alpha,\sigma}({\bf k}).\label{eq:CM2}
\end{equation}
The Fourier transform of $[C]_{{\bf r}\sigma,{\bf r}'\sigma'}$ is
given by $[C({\bf k})]_{\sigma\sigma'}=\sum_{{\bf r},{\bf r}'}[C]_{{\bf r}\sigma,{\bf r}'\sigma'}e^{-i{\bf k}\cdot({\bf r}-{\bf r}')}$.
Applying this transformation to Eq.~(\ref{eq:CM2}), we find
\begin{equation}
	[C({\bf k})]_{\sigma\sigma'}=\sum_{\alpha=1}^{2}u_{\alpha,\sigma}({\bf k})u_{\alpha,\sigma'}({\bf k})=[P_{{\rm occ}}({\bf k})]_{\sigma\sigma'}.\label{eq:Ck}
\end{equation}
Under PBC, the further calculation of $C_{{\rm A}}$ is determined
by how we choose the subsystem A. For example, if we take a bipartition
at $n=N_{y}/2$ while maintaining the translational invariance of
the system along the $x$ direction, the quasimomentum $k_{x}$ is
conserved. In this case, we can consider the inverse Fourier transformation
of $C({\bf k})$ along $y$ direction, given by 
\begin{equation}
	[C(k_{x})]_{n\sigma,n'\sigma'}=\frac{1}{N_{y}}\sum_{k_{y}\in{\rm BZ}}e^{ik_{y}(n-n')}[P_{{\rm occ}}({\bf k})]_{\sigma\sigma'}.\label{eq:Ckx}
\end{equation}
Identifying the subsystem A with all the unit cells ${\bf r}=(m,n)$
having $n\in[1,N_{y}/2]$, we can obtain the correlation matrix $C_{{\rm A}}(k_{x})$
of subsystem A in Eq.~(\ref{eq:Ckx}) as the block of $C(k_{x})$
with $1\leq n,n'\leq N_{y}/2$ for all $\sigma,\sigma'$. Another
typical bipartition can be taken at $m=N_{x}/2$ for all $n$, which
maintains the translational invariance of the system along the $y$
direction under PBC. In this case, the $k_{y}$-parameterized correlation
matrix takes the form
\begin{equation}
	[C(k_{y})]_{m\sigma,m'\sigma'}=\frac{1}{N_{x}}\sum_{k_{x}\in{\rm BZ}}e^{ik_{x}(m-m')}[P_{{\rm occ}}({\bf k})]_{\sigma\sigma'}.\label{eq:Cky}
\end{equation}
Let the subsystem A to comprise all unit cells with $m\in[1,N_{x}/2]$,
we can also identify the correlation matrix $C_{{\rm A}}(k_{y})$
of subsystem A in Eq.~(\ref{eq:Cky}) as the block of $C(k_{y})$
with $1\leq m,m'\leq N_{x}/2$ for all $\sigma,\sigma'$. Due to the
transposition-symmetric structure of our Hamiltonian $H({\bf k})$,
the $C_{{\rm A}}(k_{x})$ and $C_{{\rm A}}(k_{y})$ are expected to
possess consistent eigenspectrum in momentum space. Therefore, without
losing generality, we will focus on the ES of $C_{{\rm A}}(k_{x})$
in our later numerical studies.

Finally, to reveal the edge states and bulk-boundary correspondence,
we need to investigate the eigensystem of our model under open boundary
conditions (OBC). This can be achieved by first applying a Fourier
transformation to the Hamiltonian matrix $H({\bf k})$, and then fix
the boundary condition appropriately. For example, if we take the
PBC and OBC along the $x$ and $y$ directions, respectively, the
Hamiltonian matrix $H(k_{x})$ obtained after Fourier transforming
$H({\bf k})$ to real space along the $y$ direction reads
\begin{equation}
	[H(k_{x})]_{n\sigma,n'\sigma'}=\frac{1}{N_{y}}\sum_{k_{y}\in{\rm BZ}}e^{ik_{y}(n-n')}[H({\bf k})]_{\sigma\sigma'},\label{eq:Hkx}
\end{equation}
where $\sigma,\sigma'=1,2,3$. $[H({\bf k})]_{\sigma\sigma'}$ is
the matrix element of $H({\bf k})$ with row index $\sigma$ and column
index $\sigma'$. Truncating the unit cell indices along the $y$
direction to a finite range $n,n'=1,...,N_{y}$ then leads to a lattice
with boundaries open at ${\bf r}=(m,1)$ and ${\bf r}=(m,N_{y})$
for all $m$. Similarly, if we take the PBC and OBC along the $y$
and $x$ directions, the Hamiltonian matrix $H(k_{y})$ obtained after
Fourier transforming $H({\bf k})$ to real space along the $x$ direction
is given by 
\begin{equation}
	[H(k_{y})]_{m\sigma,m'\sigma'}=\frac{1}{N_{x}}\sum_{k_{x}\in{\rm BZ}}e^{ik_{x}(m-m')}[H({\bf k})]_{\sigma\sigma'}.\label{eq:Hky}
\end{equation}
Truncating the unit cell indices along the $x$ direction to a finite
range $m,m'=1,...,N_{x}$ then leads to a lattice with boundaries
open at ${\bf r}=(1,n)$ and ${\bf r}=(N_{x},n)$ for all $n$. In
our explicit model investigations, we will focus on the spectrum of
$H(k_{x})$ to understand the edge states and bulk-boundary correspondence.
The spectrum of $H(k_{y})$ can be obtained similarly to generate
consistent results.

Overall, we have introduced a toolkit to describe and characterize
nontrivial Euler topology in non-Hermitian gapped systems. In the
next section, we will uncover the presence of non-Hermitian topological
Euler insulator phases by exploring a series of representative models,
all sharing the same construction recipe with the general Hamiltonian
$H({\bf k})$ in Eq.~(\ref{eq:Hk}).

\section{Results\label{sec:Res}}

In this section, we apply the theoretical framework introduced in
the last section to three typical models of non-Hermitian Euler insulators.
The generating vectors of these models originate from the 2D Qi-Wu-Zhang
(QWZ) model, a generalized QWZ model with second-neighbor hoppings,
and a square-lattice realization of Haldane Chern insulator model.
In each case, we will demonstrate the physics of Euler topology from
a unified perspective of bulk gap structures, topological invariants,
entanglement spectrum and bulk-boundary correspondence. The results
illustrate that rich phases and transitions of topological Euler insulators
could appear in non-Hermitian systems.

\subsection{Model I\label{subsec:M1}}

To explicitly reveal non-Hermitian topological phases characterized
by the Euler class, we first consider a model with the generating
vector ${\bf h}_{{\rm I}}({\bf k})=(h_{{\rm I}x},h_{{\rm I}y},h_{{\rm I}z})$,
where 
\begin{alignat}{1}
	h_{{\rm I}x}&=  \sin k_{x},\label{eq:hx1}\\
	h_{{\rm I}y}&=  \sin k_{y},\label{eq:hy1}\\
	h_{{\rm I}z}&=  \cos k_{x}+\cos k_{y}-\mu.\label{eq:hz1}
\end{alignat}
The quasimomentum ${\bf k}$ is defined in the first BZ and ${\bf k}=(k_{x},k_{y})\in[-\pi,\pi)\times[-\pi,\pi)$.
Non-Hermitian effects are introduced by letting the dimensionless
parameter $\mu$ to take complex values, i.e., $\mu=\mu_{R}+i\mu_{I}$,
where $\mu_{R}\in\mathbb{R}$ and $\mu_{I}\in\mathbb{R}$. Note in
passing that the two-band Hamiltonian ${\bf h}_{{\rm I}}({\bf k})\cdot\boldsymbol{\sigma}$,
with $\boldsymbol{\sigma}=(\sigma_{x},\sigma_{y},\sigma_{z})$, gives
the standard Bloch Hamiltonian of QWZ Chern insulator model \cite{QWZ}
in the Hermitian limit $\mu_{I}=0$.

With the generating vector ${\bf h}_{{\rm I}}$, we can construct
a three-band, non-Hermitian Euler insulator model according to the
Eq.~(\ref{eq:Hk}) as 
\begin{equation}
	H_{{\rm I}}({\bf k})=[{\bf h}_{{\rm I}}({\bf k})\cdot{\bf h}_{{\rm I}}({\bf k})]|u_{3}({\bf k})\rangle\langle\tilde{u}_{3}({\bf k})|,\label{eq:Hk1}
\end{equation}
where $|u_{3}({\bf k})\rangle={\bf h}_{{\rm I}}^{\top}({\bf k})/\sqrt{{\bf h}_{{\rm I}}({\bf k})\cdot{\bf h}_{{\rm I}}({\bf k})}$
and $\langle\tilde{u}_{3}({\bf k})|={\bf h}_{{\rm I}}({\bf k})/\sqrt{{\bf h}_{{\rm I}}({\bf k})\cdot{\bf h}_{{\rm I}}({\bf k})}$
are the corresponding left and right eigenvectors of $H_{{\rm I}}({\bf k})$
with the eigenenergy
\begin{equation}
	E_{3}({\bf k})={\bf h}_{{\rm I}}({\bf k})\cdot{\bf h}_{{\rm I}}({\bf k}).\label{eq:Ek1}
\end{equation}
The other two bands of $H({\bf k})$ are flat and degenerate, with
energies $E_{1}({\bf k})=E_{2}({\bf k})=0$ at all ${\bf k}\in{\rm BZ}$.
Their left and right eigenvectors are obtained by plugging Eqs.~(\ref{eq:hx1})--(\ref{eq:hz1})
into the Eqs.~(\ref{eq:u1k}) and (\ref{eq:u2k}). 

The spectrum gap of $H_{{\rm I}}({\bf k})$ closes when the band $E_{3}({\bf k})$
touches the other two bands $E_{1,2}({\bf k})$ at zero energy. Using
the spectral gap function $\Delta$ in Eq.~(\ref{eq:GF}), we could
identify the gapless regime of the system as the parameter domain
of $(\mu_{R},\mu_{I})$ with $\Delta=0$. Using the explicit expression
of ${\bf h}_{{\rm I}}({\bf k})$, we find the condition $|E_{3}({\bf k})|=0$
to hold if the following two equations are satisfied simultaneously,
i.e., 
\begin{alignat}{1}
	\sin^{2}k_{x}+\sin^{2}k_{y}&=  \mu_{I}^{2},\nonumber \\
	\cos k_{x}+\cos k_{y}&=  \mu_{R}.\label{eq:PB1}
\end{alignat}
Eliminating the quasimomenta $k_{x}$ and $k_{y}$, we arrive at the
parameter regime in which $\Delta=0$. It is restricted by the inequalities
\begin{alignat}{1}
	(\mu_{R}+1)^{2}+\mu_{I}^{2}&\geq  1,\nonumber \\
	(\mu_{R}-1)^{2}+\mu_{I}^{2}&\geq  1,\label{eq:PB2}\\
	\mu_{R}^{2}+2\mu_{I}^{2}&\leq  4.\nonumber 
\end{alignat}
In general, these inequalities generate a 2D parameter regime on the
$\mu_{R}$-$\mu_{I}$ plane. It is sandwiched between two inner circles
$(\mu_{R}\pm1)^{2}+\mu_{I}^{2}=1$ and an outer ellipse $\mu_{R}^{2}+2\mu_{I}^{2}=4$,
which form the phase boundaries. In the Hermitian limit $\mu_{I}=0$,
the gapless region reduces to three critical points at $\mu_{R}=0,\pm2$
\cite{TEI08}. Therefore, we expect non-Hermitian effects to be able
to reshape and greatly enrich the phases and transitions of Euler
band topology in our Model I.

\begin{figure}
	\begin{centering}
		\includegraphics[scale=0.5]{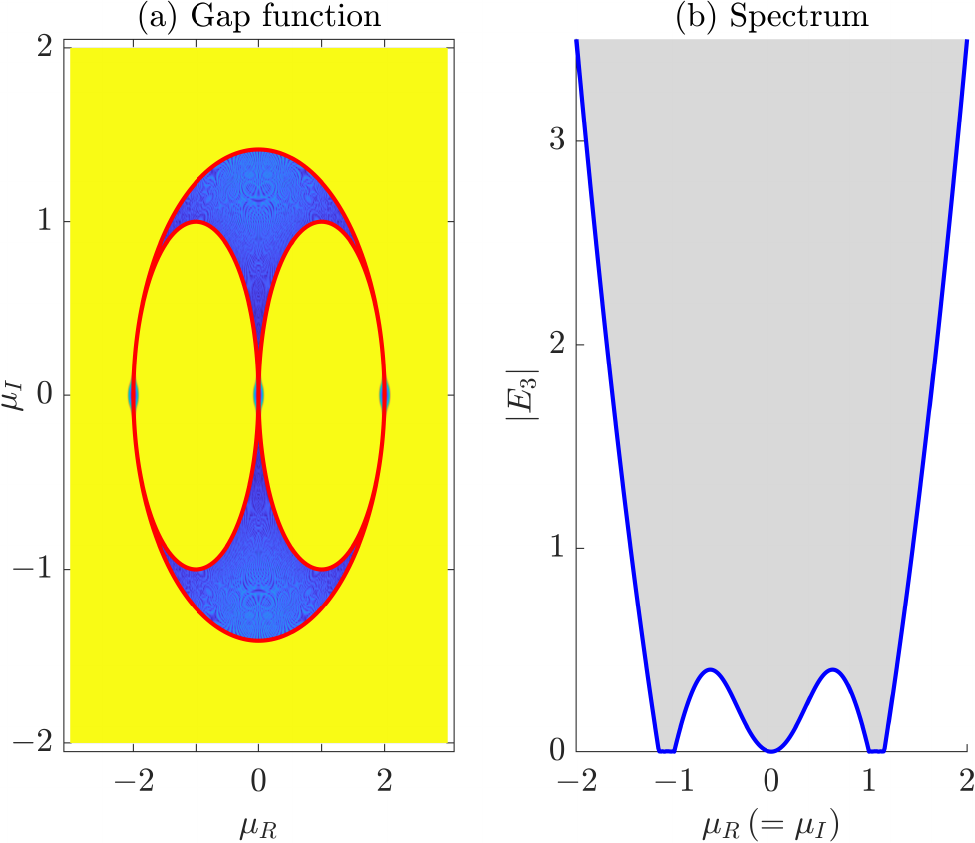}
		\par\end{centering}
	\caption{Spectral information of the Model I. (a) shows the spectral gap function
		$\Delta$ vs the real and imaginary parts of system parameter $\mu=\mu_{R}+i\mu_{I}$.
		The value of $\Delta$ is truncated at $0.01$, so that all regions
		with $\Delta\protect\geq0.01$ are colored in yellow. The red solid
		lines are phase boundaries obtained from Eq.~(\ref{eq:PB2}). (b)
		shows the spectrum of the system along the parameter line $\mu_{I}=\mu_{R}$
		in (a) under PBC. It is zoomed in around $|E_{3}|=0$ to show the
		gapped and gapless regions clearly. The gray area and blue solid line
		denote the bulk and lower edge of the highest band $E_{3}$. Two other
		flat bands $E_{1}$ and $E_{2}$ at zero energy are not shown.\label{fig:QWZspec}}
\end{figure}

In Fig.~\ref{fig:QWZspec}, we present the gap function $\Delta$
{[}Eq.~(\ref{eq:GF}){]} and the bulk spectrum $|E_{3}({\bf k})|$
{[}Eq.~(\ref{eq:Ek1}){]} of $H_{{\rm I}}({\bf k})$ under PBC. In
Fig.~\ref{fig:QWZspec}(a), the gapped and gapless regions of the
spectrum are shown in blue and yellow colors. We observe that the
gapless regime is indeed determined by the Eq.~(\ref{eq:PB2}), forming
a finite area bounded by the analytically obtained phase boundaries
(red solid lines). As expected, this critical phase reduces to three
critical points when $\mu_{I}=0$. Different from the Hermitian case,
the transitions among distinct gapped Euler insulator phases in our
system are mediated by gapless ones. This point is further illustrated
in Fig.~\ref{fig:QWZspec}(b), where the absolute spectrum $|E_{3}|$
along the parameter line $\mu_{I}=\mu_{R}$ in Fig.~\ref{fig:QWZspec}(a)
is show. We notice that despite at $\mu_{I}=0$, the band $E_{3}({\bf k})$
could touch with the two other bands at zero energy along two finite
parameter segments, which correspond to the gapless regions separating
distinct insulator phases. We will next conduct the topological characterization
of the gapped phases identified in Fig.~\ref{fig:QWZspec}.

\begin{figure}
	\begin{centering}
		\includegraphics[scale=0.5]{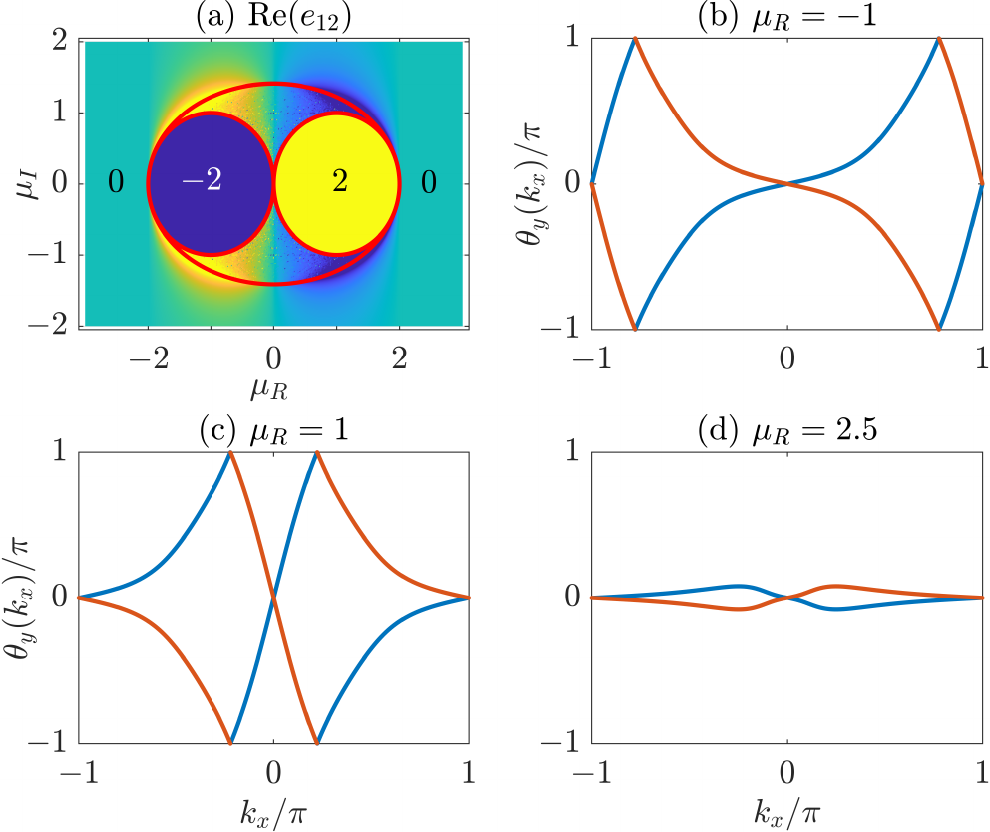}
		\par\end{centering}
	\caption{Topological characterization of the Model I. (a) shows the real parts
		of Euler class $e_{12}$ at different $\mu=\mu_{R}+i\mu_{I}$ {[}see
		Fig.~\ref{fig:ImagEC}(a) for the imaginary parts{]}. The values of
		$e_{12}$ in different gapped phases are given explicitly. The red
		solid lines represent the analytically obtained phase boundaries in
		Fig.~\ref{fig:QWZspec}. (b)--(d) show the Wilson loop spectrum $\theta_{y}(k_{x})$
		of the two degenerate bulk bands $E_{1,2}$ vs $k_{x}$ for $\mu_{I}=0.5$.
		\label{fig:QWZtopo}}
\end{figure}

Replacing the ${\bf h}$ in Eq.~(\ref{eq:PI}) by the generating vector
${\bf h}_{{\rm I}}$ with the components in Eqs.~(\ref{eq:hx1})--(\ref{eq:hz1}),
we arrive at the topological phase diagram of our Model I, as shown
in Fig.~\ref{fig:QWZtopo}(a). We find three topologically distinct
gapped phases characterized by the Euler class $e_{12}$. Two of them
are topologically nontrivial with $e_{12}=\pm2$, while the other
one is trivial with $e_{12}=0$. We refer to these gapped phases as
non-Hermitian Euler insulators. In comparison to the Hermitian case
($\mu_{I}=0$), we see that the topological phase diagram is indeed
markedly enriched, with nontrivial Euler insulator phases emerging
at finite non-Hermitian strengths. Moreover, unique transitions from
topologically nontrivial to trivial Euler insulator phases, achieved
via forming gapless intermediate phases, can be induced by increasing
the non-Hermitian parameter $\mu_{I}$. Meanwhile, we notice that
the Euler class $e_{12}$ becomes non-quantized in the gapless regime
and along the phase boundaries shown in Fig.~\ref{fig:QWZspec}(a).
In these regions, the $e_{12}$ in Eq.~(\ref{eq:PI}) becomes ill-defined
due to the vanishing denominator at gapless quasimomenta in the integrand,
making the Euler class an unreliable topological index. Therefore,
we conclude that for both Hermitian and non-Hermitian three-band systems,
the Euler class remains to be a good invariant of Euler band topology
if one of the energy bands is separated from two other degenerate
bands by a finite spectral gap.

Despite the Euler class, we also evaluate the Wilson loop spectrum
of the $P(k_{x})$ in Eq.~(\ref{eq:P}), and report the phase bands
$\theta_{y}^{\pm}(k_{x})$ for three gapped non-Hermitian phases in
Figs.~\ref{fig:QWZtopo}(b)--\ref{fig:QWZtopo}(d). In the two nontrivial
Euler insulator phases, we observe the Wilson loop winding numbers
$|w_{y}^{\pm}|=2$ for the two degenerate bands. In the trivial insulator
phase, we instead find no spectrum windings and thus $|w_{y}^{\pm}|=0$.
These results are all consistent with the obtained Euler class in
these phases according to the relationship $|e_{12}|=|w_{y}^{\pm}|$.
Therefore, the Wilson loop spectrum indeed offers a topologically
equivalent and geometrically complementary characterization for non-Hermitian
Euler topology in our system.

\begin{figure}
	\begin{centering}
		\includegraphics[scale=0.5]{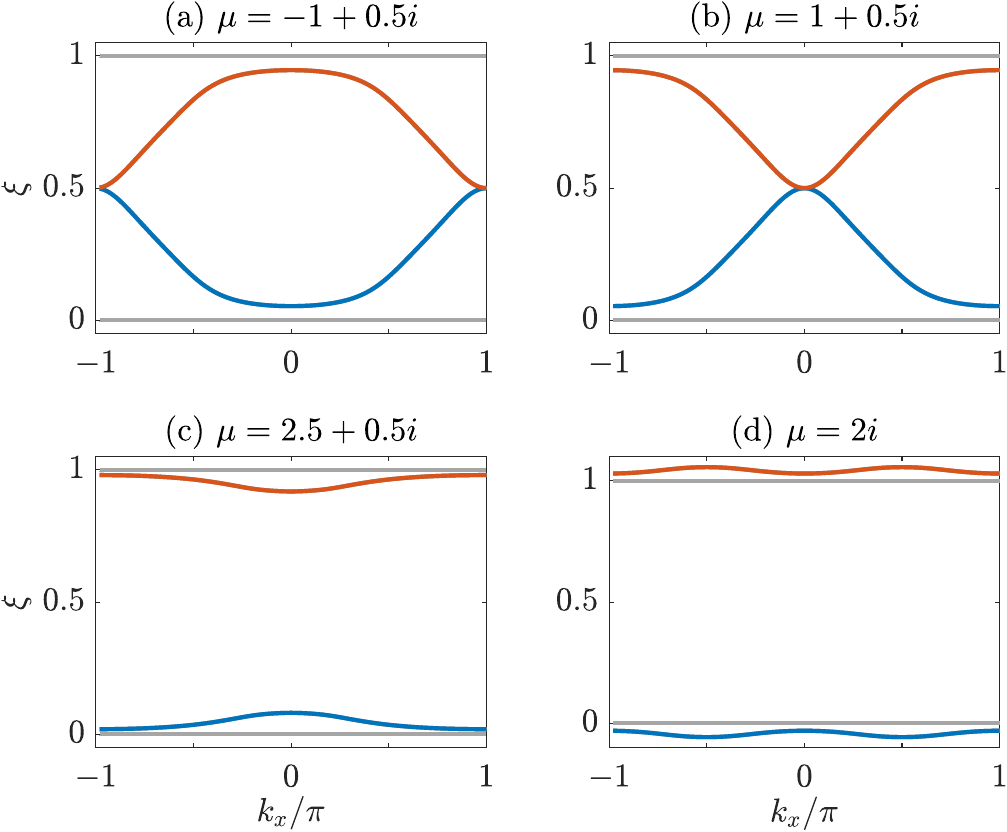}
		\par\end{centering}
	\caption{Entanglement spectrum of the Model I. The system has $N_{y}=100$
		unit cells along the $y$ direction, and an equal bipartition is taken
		along $y$ for evaluating the entanglement spectrum in each panel.
		In all panels, the gray (blue and red) lines are entanglement spectrum
		corresponding to bulk states (edge states generated along the entanglement
		cuts). \label{fig:QWZes}}
\end{figure}

To gain insights about the edge states and bulk-boundary correspondence
of non-Hermitian topological Euler insulators, we compute the spectrum
of correlation matrix $C(k_{x})$ in Eq.~(\ref{eq:Ckx}), and report
the ES for several typical cases of our Model I in Fig.~\ref{fig:QWZes}.
In each panel, the gray lines at $\xi=0,1$ are associated with bulk
bands, while the blue and red lines correspond to edge bands forming
along the $x$-directed entanglement cuts at $n=1$ and $n=N_{y}/2$.
Each branch of the edge bands is twofold degenerate. We find that
in topologically nontrivial Euler insulator phases, the two sets of
edge bands meet with each other at $\xi=0.5$, forming quadratic touching
points at $k_{x}=\pi$ and $k_{x}=0$ in Figs.~\ref{fig:QWZes}(a)
and \ref{fig:QWZes}(b). The formation of such quadratic edge-band
touchings distinguishes non-Hermitian topological Euler insulators
from well-known chiral topological phases such as Chern insulators,
in which the edge bands form linear crossings in ES. In trivial insulator
phases, the ES is instead fully gapped at $\xi=0.5$, as shown in
Figs.~\ref{fig:QWZes}(c) and \ref{fig:QWZes}(d). Notably, in the
case with a purely imaginary $\mu$, the ES could go beyond the range
$[0,1]$ of Hermitian systems, which is unique to non-Hermitian situations
as shown in Fig.~\ref{fig:QWZes}(d). Overall, we conclude that the
ES could provide clear edge-state signatures for us to distinguish
non-Hermitian Euler insulators with topologically trivial and nontrivial
bulks.

\begin{figure}
	\begin{centering}
		\includegraphics[scale=0.5]{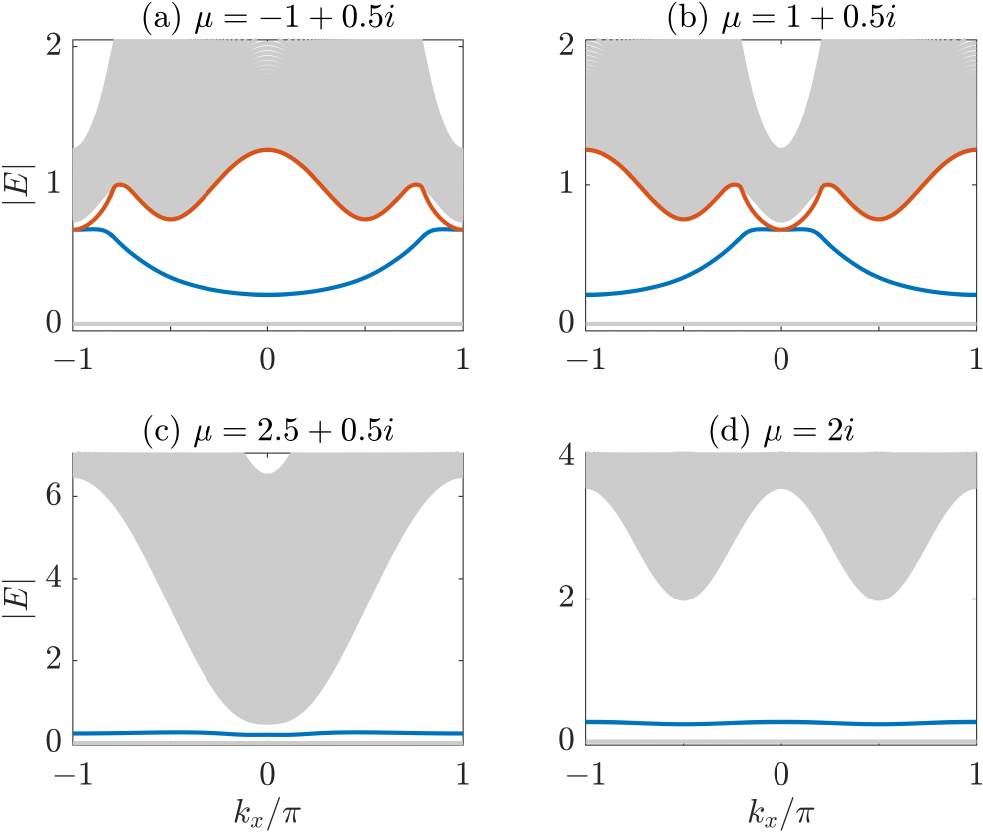}
		\par\end{centering}
	\caption{Bulk-boundary correspondence of the Model I. (a)--(d) show the absolute
		value of spectrum $|E|$ with periodic and open boundary conditions
		taken along $x$ and $y$ directions, respectively. The system has
		$N_{y}=200$ unit cells along the $y$ direction. In each panel, the
		gray (blue and red) lines correspond to the spectra of bulk (edge)
		states. \label{fig:QWZbbc}}
\end{figure}

To further reveal the bulk-boundary correspondence, we numerically
calculate the spectrum of our Model I under the PBC and OBC along
$x$ and $y$ directions. This is done by replacing the $H({\bf k})$
on the right hand side of Eq.~(\ref{eq:Hkx}) with the $H_{{\rm I}}({\bf k})$
in Eq.~(\ref{eq:Hk1}), and truncating the matrix $H(k_{x})$ at a
finite $n=N_{y}$. The resulting absolute spectra are shown in 
Fig.~\ref{fig:QWZbbc}, with the gray lines (blue and red lines) highlighting
the bulk (edge) bands. In gapped phases with nontrivial Euler band
topology, we observe two sets of edge bands with quadratic touchings
at $k_{x}=\pi$ and $k_{x}=0$ in Figs.~\ref{fig:QWZbbc}(a) and \ref{fig:QWZbbc}(b),
respectively. They are coincident with the quadratic touchings observed
in the corresponding ES in Figs.~\ref{fig:QWZes}(a) and \ref{fig:QWZes}(b),
thus forming decisive signatures of bulk-boundary correspondence for
non-Hermitian topological Euler insulators. In trivial gapped phases,
we only observe a single branch of twofold degenerate edge band. It
resides fully inside the spectral gap between the two flat bands at
zero energy and the higher dispersive band, as shown in Figs.~\ref{fig:QWZbbc}(c)
and \ref{fig:QWZbbc}(d). This is also consistent with the ES in 
Figs.~\ref{fig:QWZes}(c) and \ref{fig:QWZes}(d), where no touchings between
edge bands are observed. 

To sum up, we find that our Model I possesses rich phases and transitions
induced by non-Hermitian effects. The gapped insulating phases can
be topologically characterized by the Euler class and the Wilson loop
winding number. Two non-Hermitian topological Euler insulator phases
with $e_{12}=\pm2$ are identified. They further support degenerate
edge bands with quadratic touchings inside the gap of energy and entanglement
spectra, thereby establishing a topological bulk-boundary correspondence.
In the next subsection, we will show that phases and transitions incorporating
non-Hermitian Euler topology with larger Euler class can also be realized
following our general construction.

\subsection{Model II\label{subsec:M2}}

We next consider a model of non-Hermitian Euler insulator that can
show a large Euler class. This model originates from the generating
vector ${\bf h}_{{\rm II}}({\bf k})=(h_{{\rm II}x},h_{{\rm II}y},h_{{\rm II}z})$
of a QWZ model with next-nearest-neighbor hoppings \cite{QWZNNN},
where
\begin{alignat}{1}
	h_{{\rm II}x}&=  \sin k_{x},\label{eq:hx2}\\
	h_{{\rm II}y}&=  \sin k_{y},\label{eq:hy2}\\
	h_{{\rm II}z}&=  \cos k_{x}+\cos k_{y}+\lambda\cos(k_{x}+k_{y})-\mu.\label{eq:hz2}
\end{alignat}
We still introduce non-Hermitian effects by setting the dimensionless
parameter $\mu$ to be complex, i.e., $\mu=\mu_{R}+i\mu_{I}$, where
$\mu_{R},\mu_{I}\in\mathbb{R}$. The hopping parameter $\lambda\in\mathbb{R}$
leads to long-range couplings, and the Model I is recovered when $\lambda=0$.
We will set $\lambda=2$ throughout the following studies of this
subsection.

A three-band, non-Hermitian Euler insulator model can be constructed
by replacing the ${\bf h}({\bf k})$ in Eq.~(\ref{eq:Hk}) with ${\bf h}_{{\rm II}}({\bf k})$,
yielding 
\begin{equation}
	H_{{\rm II}}({\bf k})=[{\bf h}_{{\rm II}}({\bf k})\cdot{\bf h}_{{\rm II}}({\bf k})]|u_{3}({\bf k})\rangle\langle\tilde{u}_{3}({\bf k})|,\label{eq:Hk2}
\end{equation}
where $|u_{3}({\bf k})\rangle={\bf h}_{{\rm II}}^{\top}({\bf k})/\sqrt{{\bf h}_{{\rm II}}({\bf k})\cdot{\bf h}_{{\rm II}}({\bf k})}$
and $\langle\tilde{u}_{3}({\bf k})|={\bf h}_{{\rm II}}({\bf k})/\sqrt{{\bf h}_{{\rm II}}({\bf k})\cdot{\bf h}_{{\rm II}}({\bf k})}$
are the left and right eigenvectors of the $H_{{\rm II}}({\bf k})$
with the eigenenergy
\begin{equation}
	E_{3}({\bf k})={\bf h}_{{\rm II}}({\bf k})\cdot{\bf h}_{{\rm II}}({\bf k}).\label{eq:Ek2}
\end{equation}
The other two bands of $H_{{\rm II}}({\bf k})$ are flat and degenerate,
with energies $E_{1}({\bf k})=E_{2}({\bf k})=0$ for all ${\bf k}\in{\rm BZ}$.
Their eigenvectors are found by inserting Eqs.~(\ref{eq:hx2})--(\ref{eq:hz2})
into the Eqs.~(\ref{eq:u1k}) and (\ref{eq:u2k}).

The energy gap of $H_{{\rm II}}({\bf k})$ closes when the bands $E_{3}({\bf k})$
and $E_{1,2}({\bf k})$ are touched at zero energy. Using the gap
function $\Delta$ in Eq.~(\ref{eq:GF}), we identify the gapless
regime of the system as the parameter domain of $(\mu_{R},\mu_{I},\lambda)$
with $\Delta=0$. Using the explicit expression of ${\bf h}_{{\rm II}}({\bf k})$,
we find the condition $|E_{3}({\bf k})|=0$ to hold if the following
two equations are satisfied simultaneously, i.e., 
\begin{alignat}{1}
	\sin^{2}k_{x}+\sin^{2}k_{y}& = \mu_{I}^{2},\nonumber \\
	\cos k_{x}+\cos k_{y}+\lambda\cos(k_{x}+k_{y})& = \mu_{R}.\label{eq:PB3}
\end{alignat}
At each fixed $\lambda$, the combination of these equations will
lead to a cubic equation for the system parameters $\mu_{R}$ and
$\mu_{I}$, the explicit solutions of which are involved. Therefore,
we resort to compute the gapless regime and phase boundaries numerically for our Model II.

\begin{figure}
	\begin{centering}
		\includegraphics[scale=0.5]{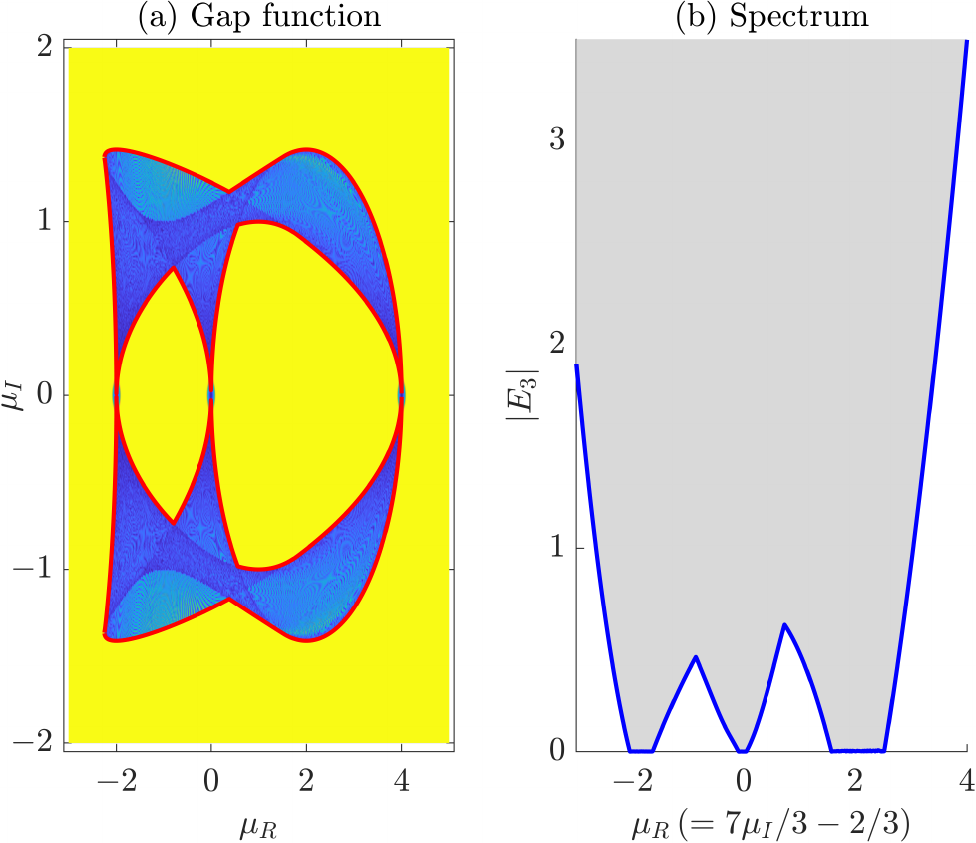}
		\par\end{centering}
	\caption{Spectral information of the Model II. The system parameter is set
		as $\lambda=2$ for both panels. (a) shows the spectral gap function
		$\Delta$ vs the real and imaginary parts of $\mu=\mu_{R}+i\mu_{I}$.
		The value of $\Delta$ is truncated at $0.01$, so that all regions
		with $\Delta\protect\geq0.01$ are colored in yellow. The red solid
		lines are phase boundaries extracted numerically from Eq.~(\ref{eq:PB3}).
		(b) shows the spectrum of the system along the parameter line $\mu_{I}=(3\mu_{R}+2)/7$
		in (a) under PBC. It is zoomed in around $|E_{3}|=0$ to show the
		gapped and gapless regions clearly. The gray area and blue solid line
		denote the bulk and lower edge of the highest band $E_{3}$. Two other
		flat bands $E_{1}$ and $E_{2}$ at zero energy are not shown. \label{fig:QWZNNNspec}}
\end{figure}

In Fig.~\ref{fig:QWZNNNspec}, we present the gap function $\Delta$
{[}Eq.~(\ref{eq:GF}){]} and the bulk spectrum $|E_{3}({\bf k})|$
{[}Eq.~(\ref{eq:Ek1}){]} of $H_{{\rm II}}({\bf k})$ with $\lambda=2$
under PBC. The numerically extracted gapless phase boundaries are
given by the red solid lines in Fig.~\ref{fig:QWZNNNspec}(a), which
enclose the gapless regime (in blue) sandwiched between gapped phases
(in yellow). In the Hermitian limit $\mu_{I}=0$, the gapless phase
shrinks to three critical points at $\mu_{R}=-2,0,4$. Similar to
the case of Model I, we realize that the phases and transitions in
our Model II are markedly altered by non-Hermitian effects, and transitions
among gapped phases are mediated by gapless ones for any nonzero $\mu_{I}$.
This point is further confirmed by the bulk spectrum computed along
the parameter line $\mu_{I}=(3\mu_{R}+2)/7$ in Fig.~\ref{fig:QWZNNNspec}(a),
as shown in Fig.~\ref{fig:QWZNNNspec}(b). There, we find four gapped
phases separated by three gapless regions in the spectrum. It will
soon become clear that these gapped domains belong to three topologically
distinct Euler insulator phases, which are characterized by different
Euler class and Wilson loop winding numbers.

\begin{figure}
	\begin{centering}
		\includegraphics[scale=0.5]{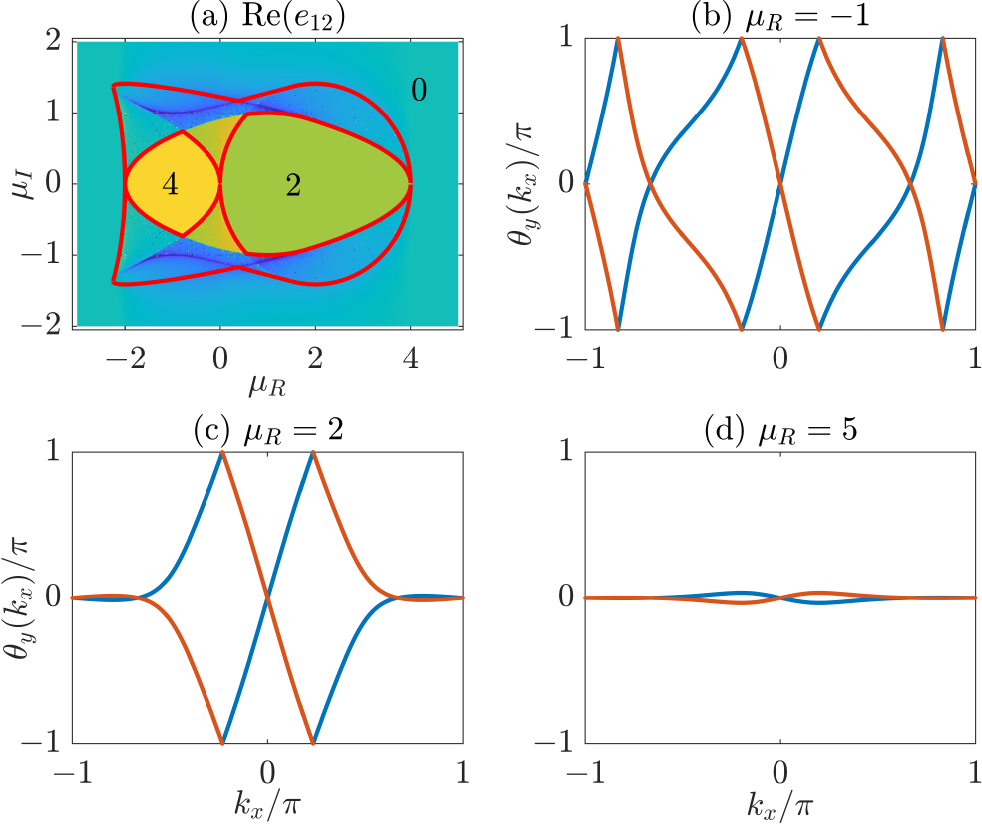}
		\par\end{centering}
	\caption{Topological characterization of the Model II. (a) shows the real parts
		of Euler class $e_{12}$ at different $\mu=\mu_{R}+i\mu_{I}$ {[}see
		Fig.~\ref{fig:ImagEC}(b) for the imaginary parts{]} for $\lambda=2$.
		The values of $e_{12}$ in different gapped phases are given explicitly.
		The red solid lines represent the numerically extracted phase boundaries
		in Fig.~\ref{fig:QWZNNNspec}(a). (b)--(d) show the Wilson loop spectrum
		$\theta_{y}(k_{x})$ of two degenerate bulk bands $E_{1,2}$ vs $k_{x}$
		for $\mu_{I}=0.5$. \label{fig:QWZNNNtopo}}
\end{figure}

To reveal the Euler band topology of our non-Hermitian Model II, we
evaluate its Euler class and Wilson loop spectrum following the 
Eqs.~(\ref{eq:PI}) and (\ref{eq:Wy})--(\ref{eq:P}). The results are
presented in Fig.~\ref{fig:QWZNNNtopo}. In the phase diagram 
Fig.~\ref{fig:QWZNNNtopo}(a), we identify three gapped phases with the
Euler class $e_{12}=4,2,0$. The former two represent topologically
nontrivial non-Hermitian Euler insulators. For any nonzero $\mu_{I}$,
the transitions among distinct Euler insulator phases are mediated
by gapless intermediate phases instead of isolated critical points,
which unveils an important difference caused by non-Hermitian effects.
Notably, we observe an Euler insulator phase with $e_{12}=4$ at nonzero
$\mu_{I}$, which demonstrates that non-Hermitian Euler insulators
with the Euler class larger than $2$ could emerge due to the interplay
between long-range hoppings and non-Hermitian effects. The Wilson
loop spectra for three representative phases with the Euler class
$e_{12}=4,2,0$ in Fig.~\ref{fig:QWZNNNtopo}(a) are shown in 
Figs.~\ref{fig:QWZNNNtopo}(b)--\ref{fig:QWZNNNtopo}(d). In each case,
a simple counting of the spectral winding number leads to $|e_{12}|=|w_{y}^{\pm}|$,
which further confirms the topological aspects of our non-Hermitian
Euler bands. Putting together, we conclude that the collaboration
between long-range couplings and non-Hermitian effects could indeed
result in topological Euler insulators with richer phase diagram and
larger Euler class.

\begin{figure}
	\begin{centering}
		\includegraphics[scale=0.5]{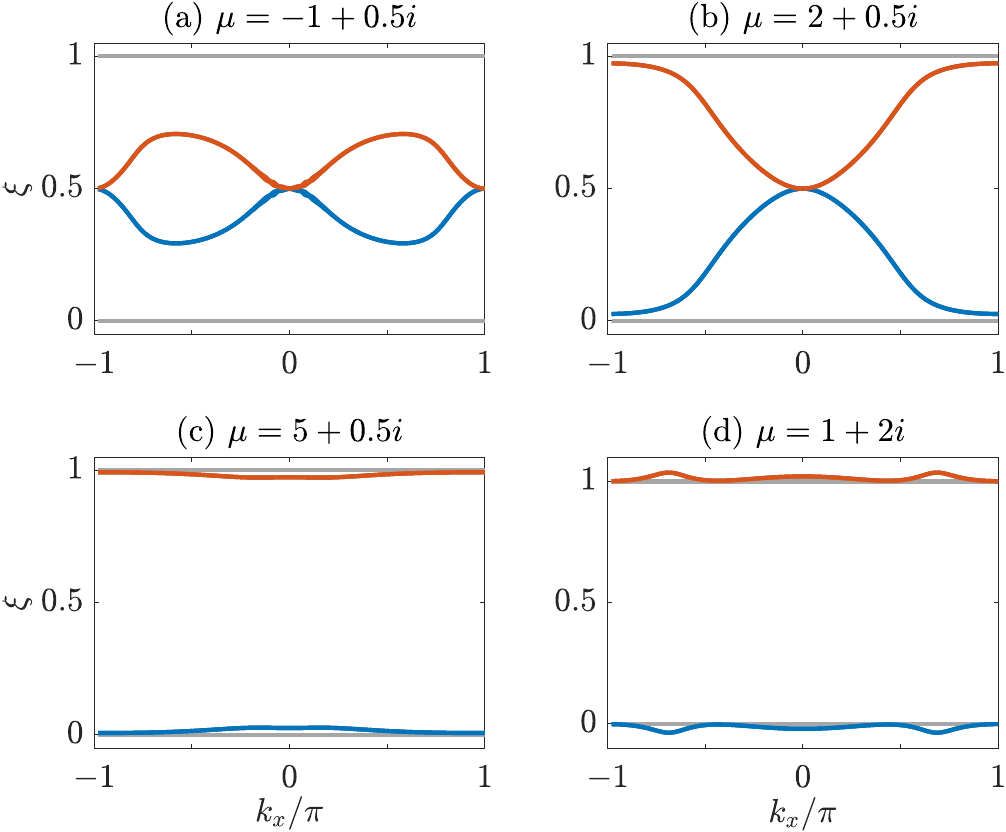}
		\par\end{centering}
	\caption{Entanglement spectrum of the Model II. The system has $N_{y}=100$
		unit cells along the $y$ direction, and an equal bipartition is taken
		along $y$ for computing the entanglement spectrum in each panel.
		In all panels, the gray (blue and red) lines are entanglement spectrum
		associated with bulk states (edge states generated along the entanglement
		cuts). We set $\lambda=2$ for all presented cases. \label{fig:QWZNNNes}}
\end{figure}

To extract the features of edge states and bulk-boundary correspondence,
we next investigate the ES of our Model II. The spectrum of correlation
matrix $C(k_{x})$ is calculated following Eq.~(\ref{eq:Ckx}), and
the ES for several typical cases of the system $H_{{\rm II}}$ are
reported in Fig.~\ref{fig:QWZNNNes}. In the case with $e_{12}=4$,
we find two quadratic touchings at $k_{x}=0,\pi$ in the branches
of ES related to edge bands along the entanglement cuts in Fig.~\ref{fig:QWZNNNes}(a).
This case is clearly different from the situations with smaller Euler
class as we encountered in Fig.~\ref{fig:QWZes}. In the case with
$e_{12}=2$, we find only one quadratic touching at $k_{x}=0$ between
the entanglement edge bands in Fig.~\ref{fig:QWZNNNes}(b). In cases
with $e_{12}=0$, the entanglement edge bands are gapped and there
are no touchings between them, as shown in Figs.~\ref{fig:QWZNNNes}(c)
and \ref{fig:QWZNNNes}(d). Note in passing that in the deep non-Hermitian
regime, the ES could go beyond the range $[0,1]$ as shown in 
Fig.~\ref{fig:QWZNNNes}(d), which is unique to non-Hermitian systems.
Overall, we identify a bulk-boundary relation between the Euler class
of non-Hermitian Bloch bands and the edge bands of ES, i.e., the absolute
value of Euler class $|e_{12}|$ corresponds to the number of quadratic
touchings between entanglement edge bands in non-Hermitian topological
Euler insulators. This relation would also help us to understand the
bulk-boundary correspondence identified in the energy spectrum. 

\begin{figure}
	\begin{centering}
		\includegraphics[scale=0.5]{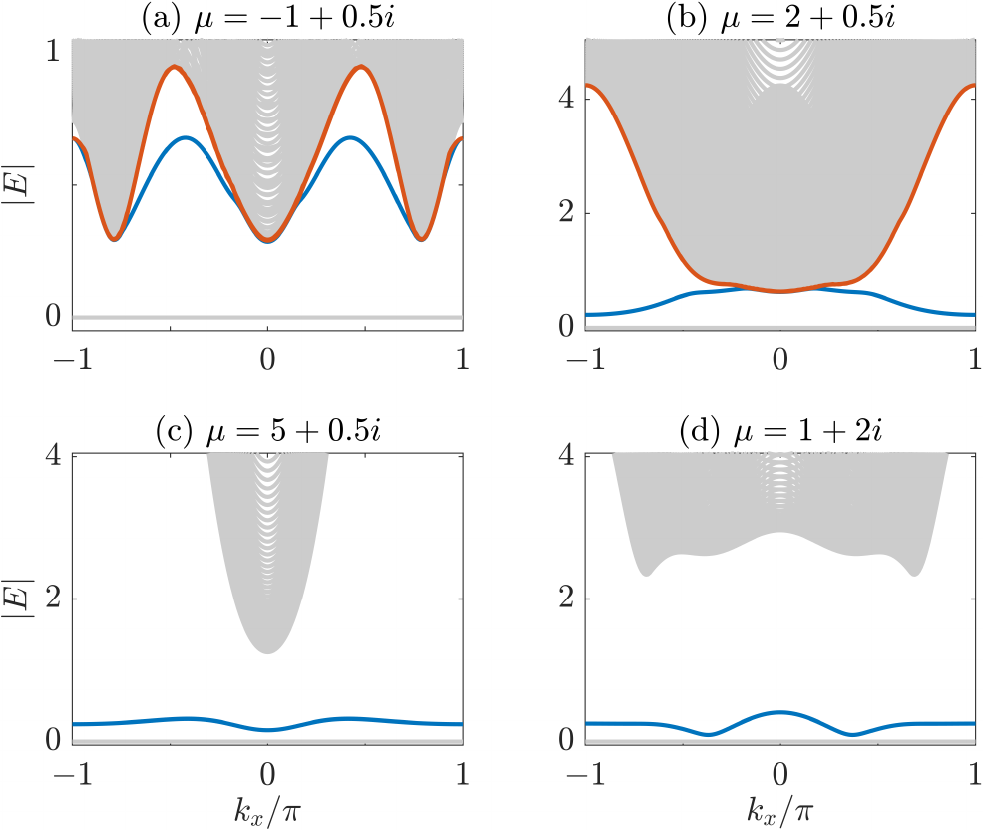}
		\par\end{centering}
	\caption{Bulk-boundary correspondence of the Model II. (a)--(d) show the absolute
		value of spectrum $|E|$ with periodic and open boundary conditions
		taken along the $x$ and $y$ directions, respectively. The system
		has $N_{y}=200$ unit cells along the $y$ direction. In each panel,
		the gray (blue and red) lines denote the spectra of bulk (edge) states.
		We set the system parameter $\lambda=2$ for all panels. \label{fig:QWZNNNbbc}}
\end{figure}

Finally, we investigate the edge states and bulk-boundary correspondence
by evaluating the spectrum of Model II under PBC and OBC along the
$x$ and $y$ directions, respectively. The results for four representative
phases are presented in Fig.~\ref{fig:QWZNNNbbc}. In the case with
$\mu=-1+0.5i$, we have the Euler class $e_{12}=4$, and the energy
spectrum of edge bands show two quadratic touchings at the high-symmetry
momentum $k_{x}=0,\pi$, as shown in Fig.~\ref{fig:QWZNNNbbc}(a).
This is consistent with the quadratic touchings we observed in the
corresponding ES of Fig.~\ref{fig:QWZNNNes}(a). In the case with
$\mu=2+0.5i$, we find the Euler class $e_{12}=2$, and the spectrum
of edge bands in Fig.~\ref{fig:QWZNNNbbc}(b) show one quadratic touching
at $k_{x}=0$, which is also coincident with the quadratic touching
observed in the ES of Fig.~\ref{fig:QWZNNNes}(b). In the other two
cases, the bulk bands are topologically trivial, and we observe no
touchings in the spectra of edge bands, as shown in Figs.~\ref{fig:QWZNNNbbc}(c)
and \ref{fig:QWZNNNbbc}(d). We conclude that in the energy spectrum
of non-Hermitian topological Euler insulators, the Euler class of
bulk bands also counts the number of quadratic touchings between edge
bands at high-symmetry momenta, thereby yielding a description of
the bulk-boundary correspondence. We notice that this correspondence
holds not only for the minimal nontrivial Euler class with $e_{12}=2$,
but also works for a larger Euler class with $e_{12}=4$. However,
for lattice models with more complicated geometry, the bulk-boundary
correspondence identified here may require a generalized interpretation,
as will be uncovered in the following subsection. 

\subsection{Model III\label{subsec:M3}}

We now investigate a non-Hermitian Euler insulator model with a different
geometric origin. The generating vector ${\bf h}_{{\rm III}}({\bf k})=(h_{{\rm III}x},h_{{\rm III}y},h_{{\rm III}z})$
of this model may be viewed as rooted in a square-lattice realization
of the Haldane honeycomb lattice model \cite{Haldane}. The components
of ${\bf h}_{{\rm III}}({\bf k})$ are given by
\begin{alignat}{1}
	h_{{\rm III}x}= & 1+\cos k_{x}+\cos(k_{x}-k_{y}),\label{eq:hx3}\\
	h_{{\rm III}y}= & \sin k_{x}+\sin(k_{x}-k_{y}),\label{eq:hy3}\\
	h_{{\rm III}z}= & \mu+2\lambda\sin\phi[\sin k_{x}-\sin k_{y}-\sin(k_{x}-k_{y})],\label{eq:hz3}
\end{alignat}
where $\lambda\in\mathbb{R}$ and $\phi\in[0,2\pi]$ correspond to
hopping and phase parameters, respectively. Non-Hermitian effects
are introduced by letting $\mu=\mu_{R}+i\mu_{I}$, where $\mu_{R},\mu_{I}\in\mathbb{R}$.
We set $\lambda=1/(3\sqrt{3})$ in the following calculations and
focus on the cases with $\phi=\pm\pi/2$. Possible phases that could
be realized at different $\mu$ for each fixed pair of parameters
$(\lambda,\phi)$ will be considered.

Replacing the ${\bf h}({\bf k})$ in Eq.~(\ref{eq:Hk}) with ${\bf h}_{{\rm III}}({\bf k})$,
we obtain a three-band model of non-Hermitian topological Euler insulator,
whose Bloch Hamiltonian reads
\begin{equation}
	H_{{\rm III}}({\bf k})=[{\bf h}_{{\rm III}}({\bf k})\cdot{\bf h}_{{\rm III}}({\bf k})]|u_{3}({\bf k})\rangle\langle\tilde{u}_{3}({\bf k})|.\label{eq:Hk3}
\end{equation}
Here, $|u_{3}({\bf k})\rangle={\bf h}_{{\rm III}}^{\top}({\bf k})/\sqrt{{\bf h}_{{\rm III}}({\bf k})\cdot{\bf h}_{{\rm III}}({\bf k})}$
and $\langle\tilde{u}_{3}({\bf k})|={\bf h}_{{\rm III}}({\bf k})/\sqrt{{\bf h}_{{\rm III}}({\bf k})\cdot{\bf h}_{{\rm III}}({\bf k})}$
are the left and right eigenvectors of $H_{{\rm III}}({\bf k})$ with
the eigenenergy
\begin{equation}
	E_{3}({\bf k})={\bf h}_{{\rm III}}({\bf k})\cdot{\bf h}_{{\rm III}}({\bf k}).\label{eq:Ek3}
\end{equation}
They other two degenerate bands of $H_{{\rm III}}({\bf k})$ are flat
with the energies $E_{1}({\bf k})=E_{2}({\bf k})=0$ for all ${\bf k}\in{\rm BZ}$.
Their eigenvectors are given by plugging Eqs.~(\ref{eq:hx3})--(\ref{eq:hz3})
into the Eqs.~(\ref{eq:u1k}) and (\ref{eq:u2k}).

The spectral gap of $H_{{\rm III}}({\bf k})$ vanishes when the bands
$E_{3}({\bf k})$ and $E_{1,2}({\bf k})$ are met at zero energy.
The gapless regime of the system is given by the parameter space $(\mu_{R},\mu_{I},\lambda,\phi)$
within which the gap function in Eq.~(\ref{eq:GF}) satisfying $\Delta=0$.
Using the explicit expression of ${\bf h}_{{\rm III}}({\bf k})$,
we find the condition $|E_{3}({\bf k})|=0$ to hold if the following
two equations are fulfilled together, i.e., 
\begin{alignat}{1}
	3+2\cos k_{x}+2\cos k_{y}+2\cos(k_{x}-k_{y})& = \mu_{I}^{2},\nonumber \\
	2\lambda\sin\phi[\sin(k_{x}-k_{y})+\sin k_{y}-\sin k_{x}]& = \mu_{R}.\label{eq:PB4}
\end{alignat}
Eliminating the quasimomenta $k_{x}$ and $k_{y}$, we can find the
parameter regime in which $\Delta=0$. It is decided by the intersection
of the following two inequalities
\begin{alignat}{1}
	\left(\frac{\mu_{R}}{\lambda\sin\phi}\right)^{2}& \geq (3+|\mu_{I}|)^{3}(1-|\mu_{I}|),\nonumber \\
	\left(\frac{\mu_{R}}{\lambda\sin\phi}\right)^{2}& \leq (3-|\mu_{I}|)^{3}(1+|\mu_{I}|).\label{eq:PB5}
\end{alignat}
In the Hermitian limit, these inequalities reduce to a single equality
$\mu_{R}=\pm3\sqrt{3}\lambda\sin\phi$, which formally reproduces
the standard phase boundary of Haldane Chern insulator model \cite{Haldane}.
We could then expect the enrichment of Euler band topology by non-Hermitian
effects for any nonzero $\mu_{I}$.

\begin{figure}
	\begin{centering}
		\includegraphics[scale=0.5]{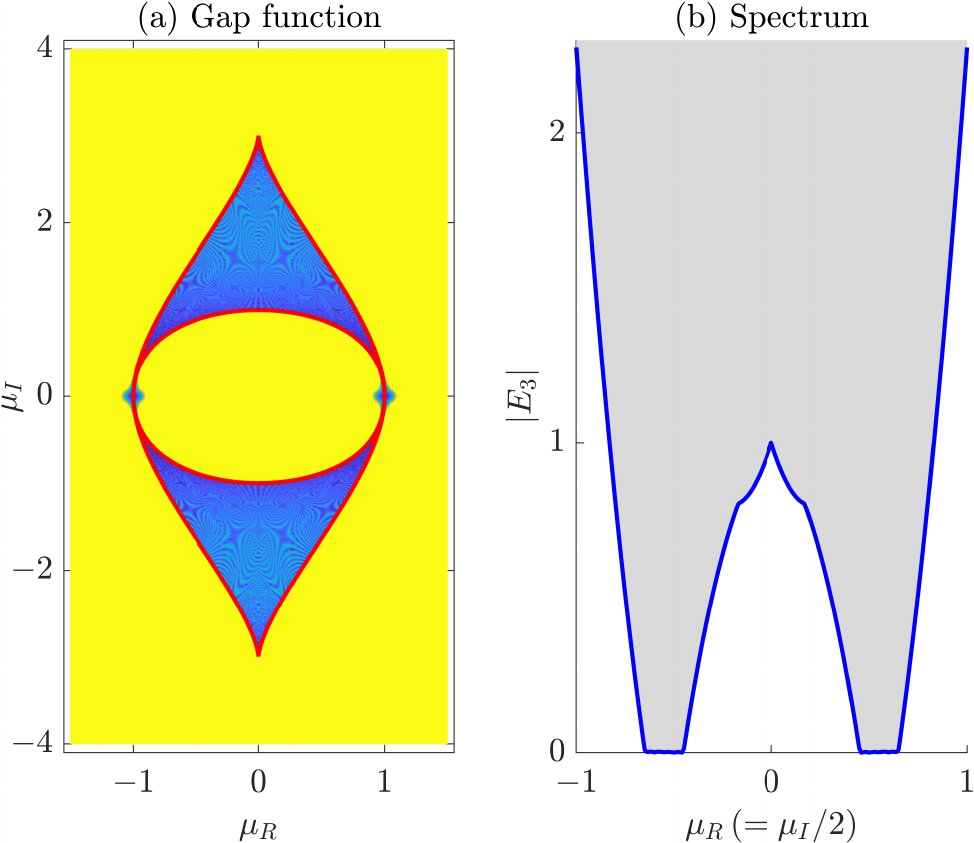}
		\par\end{centering}
	\caption{Spectral information of the Model III. The system parameters are set
		as $\phi=-\pi/2$ and $\lambda=1/\sqrt{27}$ for both panels. (a)
		shows the spectral gap function $\Delta$ vs the real and imaginary
		parts of $\mu=\mu_{R}+i\mu_{I}$. The value of $\Delta$ is truncated
		at $0.01$, so that all regions with $\Delta\protect\geq0.01$ are
		colored in yellow. The red solid lines are phase boundaries obtained
		analytically from Eq.~(\ref{eq:PB5}). (b) shows the spectrum of the
		system along the parameter line $\mu_{I}=2\mu_{R}$ in (a) under PBC.
		It is zoomed in around $|E_{3}|=0$ to show the gapped and gapless
		regions clearly. The gray area and blue solid line denote the bulk
		and lower edge of the highest band $E_{3}$. Two other flat bands
		$E_{1}$ and $E_{2}$ at zero energy are not shown. \label{fig:HDspec}}
\end{figure}

We first consider the the gap function $\Delta$ {[}Eq.~(\ref{eq:GF}){]}
and the bulk spectrum $|E_{3}({\bf k})|$ {[}Eq.~(\ref{eq:Ek3}){]}
of our Model III under PBC, with numerical results shown in Fig.~\ref{fig:HDspec}.
The red solid lines in Fig.~\ref{fig:HDspec}(a) are phase boundaries
obtained from Eq.~(\ref{eq:PB5}) after keeping only the equality
parts, i.e., $\mu_{R}^{2}=(\lambda\sin\phi)^{2}(3+|\mu_{I}|)^{3}(1-|\mu_{I}|)$
and $\mu_{R}^{2}=(\lambda\sin\phi)^{2}(3-|\mu_{I}|)^{3}(1+|\mu_{I}|)$.
We observe two disconnected gapped regions separated by a gapless
phase. The latter shrinks to two critical points at $\mu_{R}=\pm1$
in the Hermitian limit $\mu_{I}=0$ following our parameter choice.
The reconstruction of phase structure via non-Hermitian effects then
becomes clear. In Fig.~\ref{fig:HDspec}(b), we further show the bulk
spectrum of $H_{{\rm III}}({\bf k})$ along the parameter line $\mu_{I}=2\mu_{R}$
in Fig.~\ref{fig:HDspec}(a). We find three gapped regime mediated
by two gapless domains, through which phase transitions occur. Similar
to previous models, we will see that these gapped regions correspond
to topologically distinct non-Hermitian Euler insulator phases. Note
in passing that the results in Fig.~\ref{fig:HDspec} look identical
when we set the phase parameter $\phi=\pi/2$ while maintaining other
system parameters unchanged.

\begin{figure}
	\begin{centering}
		\includegraphics[scale=0.5]{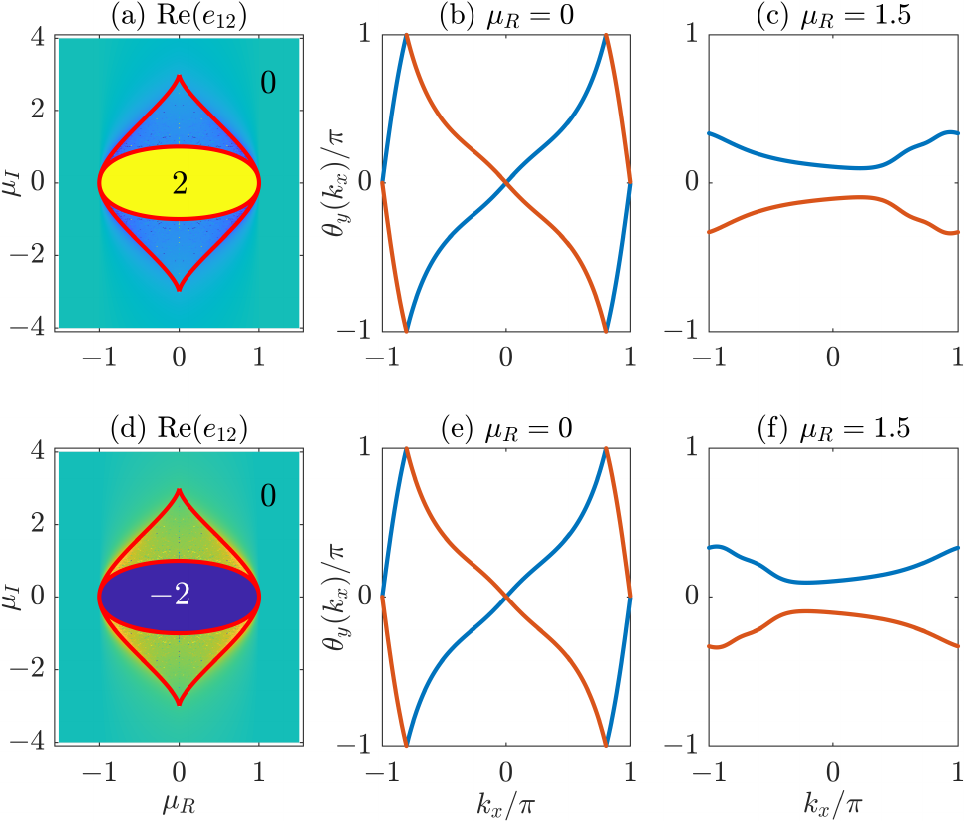}
		\par\end{centering}
	\caption{Topological characterization of the Model III. (a) and (d) show the
		real parts of Euler class $e_{12}$ at different $\mu=\mu_{R}+i\mu_{I}$
		{[}see Figs.~\ref{fig:ImagEC}(c) and \ref{fig:ImagEC}(d) for the
		imaginary parts{]} for $\phi=-\pi/2$ and $\pi/2$, respectively.
		The red solid lines represent the analytically obtained phase boundaries
		in Fig.~\ref{fig:HDspec}. The values of $e_{12}$ in different gapped
		phases are given explicitly. (b) and (c) {[}(e) and (f){]} show the
		Wilson loop spectrum $\theta_{y}(k_{x})$ of two degenerate bulk bands
		vs $k_{x}$ for $\mu_{I}=0.5$ and $\phi=-\pi/2$ {[}$\phi=\pi/2${]}.
		We set $\lambda=1/\sqrt{27}$ for all panels. \label{fig:HDtopo}}
\end{figure}

To clarify the non-Hermitian Euler topology, we compute the Euler
class and Wilson loop spectrum of our Model III following the 
Eqs.~(\ref{eq:PI}) and (\ref{eq:Wy})--(\ref{eq:P}). The resulting phase
diagrams and Wilson loop bands are shown in Fig.~\ref{fig:HDtopo}.
Including the cases with $\phi=\pm\pi/2$, we find three gapped phases
with different Euler class $e_{12}=\pm2,0$. The former two phases
are topologically nontrivial. Each of them is separated from the trivial
insulator by a gapless phase, as shown in Figs.~\ref{fig:HDtopo}(a)
and \ref{fig:HDtopo}(d), where the Euler class in Eq.~(\ref{eq:PI})
becomes ill-defined. Non-Hermitian topological Euler insulator phases
are found to survive in a finite range of $\mu_{I}$ for $\mu_{R}\in[-1,1]$,
and phase transitions between topological and trivial Euler insulator
phases occur with the increase of non-Hermitian strength $\mu_{I}$.
The topological distinction is further confirmed by the winding behavior
of Wilson loop spectrum. In the topologically nontrivial gapped phases,
we find the Wilson loop winding number $|w_{y}^{\pm}|=2$ as shown
in Figs.~\ref{fig:HDtopo}(b) and \ref{fig:HDtopo}(e). In the trivial
gapped phases, we find $|w_{y}^{\pm}|=0$ as presented in Figs.~\ref{fig:HDtopo}(c)
and \ref{fig:HDtopo}(f). They are all consistent with the topology
predicted by the Euler class in the phase diagrams. Therefore, we
conclude that following our general model construction, non-Hermitian
topological Euler insulators can be realized upon different lattice
configurations.

\begin{figure}
	\begin{centering}
		\includegraphics[scale=0.5]{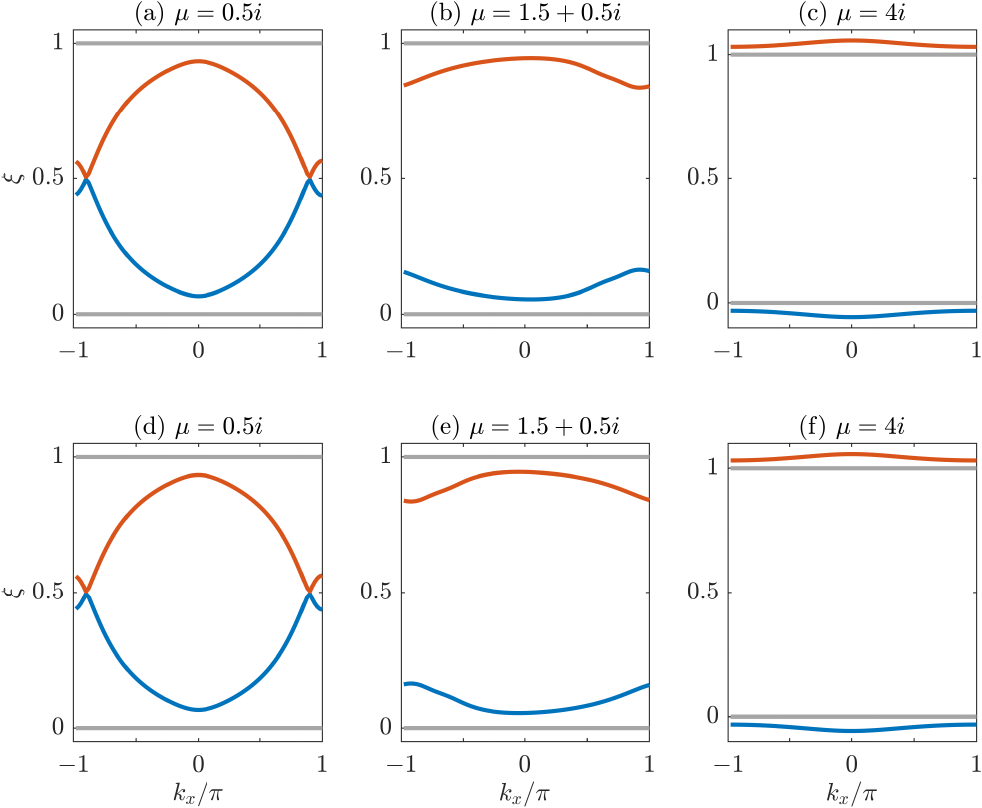}
		\par\end{centering}
	\caption{Entanglement spectrum of the Model III. The system has $N_{y}=100$
		unit cells along the $y$ direction, and an equal bipartition is taken
		along $y$ for computing the entanglement spectrum in each panel.
		Other system parameters are $(\phi,\lambda)=(-\pi/2,1/\sqrt{27})$
		for (a)--(c) and $(\phi,\lambda)=(\pi/2,1/\sqrt{27})$ for (d)--(f).
		In all panels, the gray (blue and red) lines are entanglement spectrum
		related to bulk states (edge states generated along the entanglement
		cuts). \label{fig:HDes}}
\end{figure}

We next reveal the topological bulk-boundary correspondence by considering
the ES. Similar to previous models, we compute the spectrum of correlation
matrix $C(k_{x})$ for our Model III following the Eq.~(\ref{eq:Ckx}).
The results for six typical parameter settings are shown in Fig.~\ref{fig:HDes}.
For trivial insulating phases, we indeed observe fully gapped edge
bands in the ES, as given by the blue and red curves in Figs.~\ref{fig:HDes}(b),
\ref{fig:HDes}(c), \ref{fig:HDes}(e), and \ref{fig:HDes}(f). Interestingly,
we observe in Figs.~\ref{fig:HDes}(a) and \ref{fig:HDes}(d) that
for the topologically nontrivial cases, the edge bands in ES do not
form a quadratic touching but instead developing an overlap in a finite
momentum domain. This overlapped region has two linear band crossings
at its ends and show quadratic dispersions at its top and bottom.
In the Hermitian limit $\mu_{I}=0$, it will reduce to a quadratic
touching at $k_{x}=\pi$. Therefore, the buildup of this anomalous
edge band overlap in the ES of our system has a non-Hermitian origin,
going beyond the usual expectation in Hermitian cases. To recover
the bulk-boundary correspondence identified in previous model studies,
we note that the total gradient of edge band dispersions at the two
touching points is two, which is numerically equivalent to a quadratic
touching. The two linear touchings in the entanglement edge bands
may then be traced back to the splitting of a quadratic touching point
caused by non-Hermitian effects.

\begin{figure}
	\begin{centering}
		\includegraphics[scale=0.5]{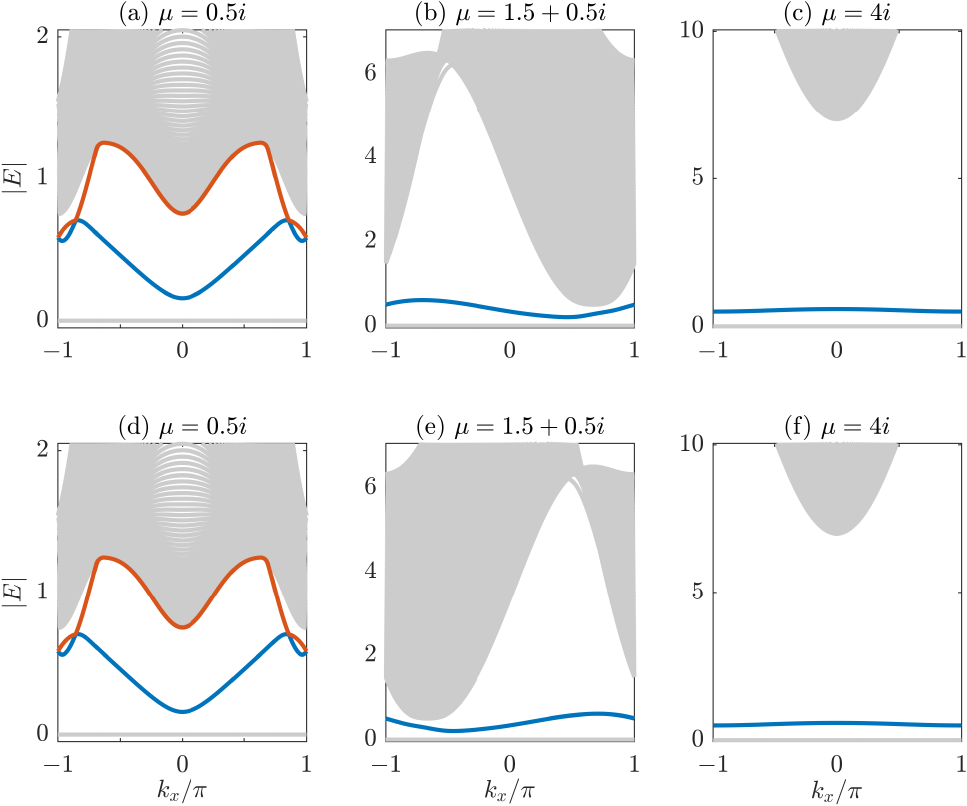}
		\par\end{centering}
	\caption{Bulk-boundary correspondence of the Model III. (a)--(c) {[}(d)--(f){]}
		show the absolute values of spectrum $|E|$ for $\phi=-\pi/2$ {[}$\phi=\pi/2${]}
		with periodic and open boundary conditions taken along the $x$ and
		$y$ directions, respectively. The system has $N_{y}=200$ unit cells
		along the $y$ direction. In each panel, the gray (blue and red) lines
		denote the spectra of bulk (edge) states. We set the system parameter
		$\lambda=1/\sqrt{27}$ for all panels. \label{fig:HDbbc}}
\end{figure}

To continue the study of edge states and bulk-boundary correspondence,
we compute the energy spectrum of our system $H_{{\rm III}}$ under
the PBC and OBC along $x$ and $y$ directions, respectively. The
numerical results are shown in Fig.~\ref{fig:HDbbc}, with the gray
(blue and red) lines represent bulk (edge) bands. In topologically
trivial phases, we find only a single branch of edge band inside the
gap between the ground state and the higher band $|E_{3}(k_{x})|$,
as shown in Figs.~\ref{fig:HDbbc}(b), \ref{fig:HDbbc}(c), \ref{fig:HDbbc}(e),
and \ref{fig:HDbbc}(f). This observation is consistent with the prediction
of the topological phase diagram and the ES at corresponding system
parameters. The nontrivial cases shown in Figs.~\ref{fig:HDbbc}(a)
and \ref{fig:HDbbc}(d) have the Euler class $e_{12}=\pm2$, and thus
representing non-Hermitian topological Euler insulators. Similar to
the ES in Figs.~\ref{fig:HDes}(a) and \ref{fig:HDes}(d), we find
an overlap regime between the two branches of edge bands, which will
shrink to a single quadratic touching at $k_{x}=\pi$ in the Hermitian
limit $\mu_{I}=0$. Therefore, the bulk-boundary correspondence revealed
in the ES is consistent with the observation in the energy spectrum.
The anomalous edge-band overlap in the spectrum of $H_{{\rm III}}$
is again triggered by non-Hermitian effects. To sum up, we find that
the topology, entanglement characteristics and bulk-boundary correspondence
of Euler bands can all be greatly reshaped by non-Hermitian effects
throughout our model studies.

\section{Discussion and conclusion\label{sec:Sum}}

\begin{figure}
	\begin{centering}
		\includegraphics[scale=0.48]{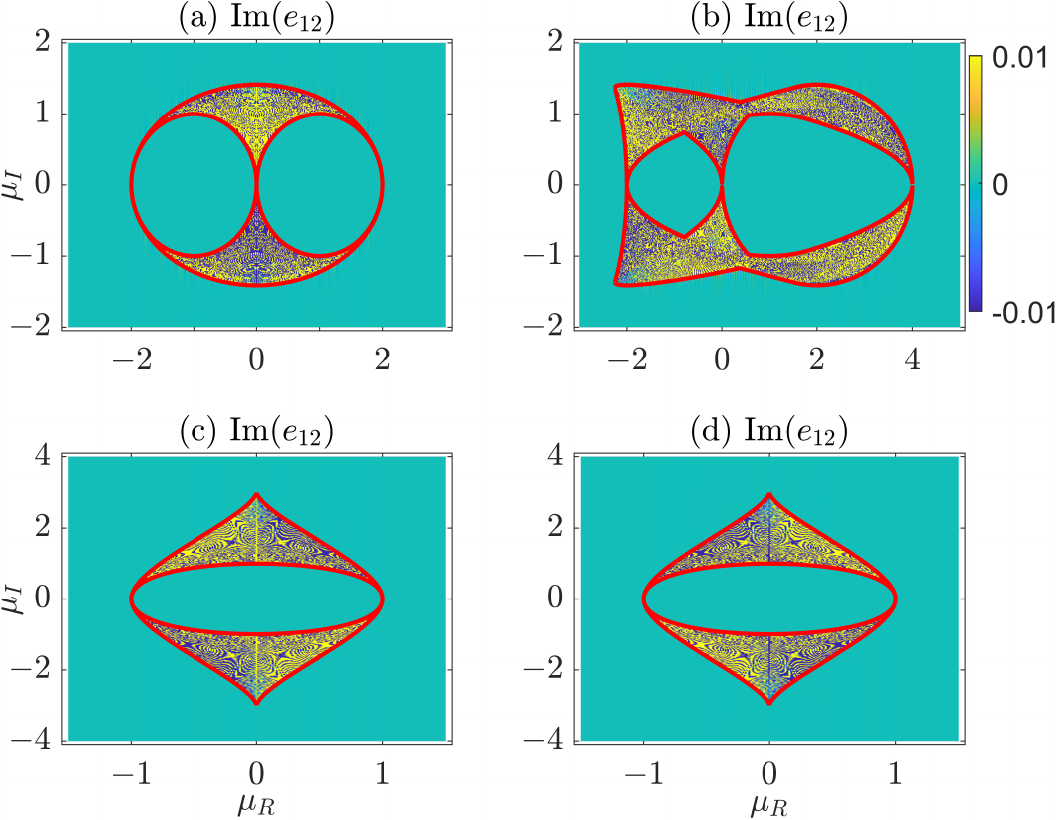}
		\par\end{centering}
	\caption{Imaginary parts of the Euler class for Model I in (a), Model II in
		(b), and Model III in (c), (d). All panels share the same color bar.
		Other system parameters are set as $\lambda=2$ for (b), $(\lambda,\phi)=(1/\sqrt{27},-\pi/2)$
		for (c) and $(\lambda,\phi)=(1/\sqrt{27},\pi/2)$ for (d). The red
		solid lines are phase boundaries obtained analytically for (a), (c),
		(d), and numerically for (b). \label{fig:ImagEC}}
\end{figure}

In the phase diagrams reported in Sec.~\ref{sec:Res} for the Models
I--III, we only show the real parts of Euler class ${\rm Re}(e_{12})$.
For each model, the Euler class $e_{12}$ is numerically obtained
from Eq.~(\ref{eq:PI}). Regarding the imaginary parts of Euler class,
we find ${\rm Im}e_{12}=0$ in all gapped regions of our considered
models, which confirms that the $e_{12}$ could indeed work as a well-defined
topological index to characterize non-Hermitian gapped phases with
nontrivial Euler topology. When the spectral gap between the band
$3$ and the two other degenerate bands at $E=0$ is closed, the $e_{12}$
becomes ill-defined. Generally, we find ${\rm Im}(e_{12})\neq0$ in
these gapless regions, as shown in Fig.~\ref{fig:ImagEC}. Meanwhile,
since the Wilson loop spectrum $\theta_{y}(k_{x})$ represents phase
bands, its winding number in momentum space is always real. We have
also verified that the winding number of Wilson loop spectrum is integer
quantized and equal in magnitudes to the Euler class in all non-Hermitian
topological Euler insulator phases we found. Nevertheless, the Wilson
loop winding number becomes ill-defined and may take non-quantized
real values in gapless phases. Therefore, neither the Euler class
nor the Wilson loop winding numbers constitute good topological indices
to characterize the non-Hermitian critical phases in our settings. Further
investigations are required to understand the presence or absence
of fully gapless non-Hermitian phases with nontrivial Euler topology.

In conclusion, we introduced a systematic approach to construct non-Hermitian
topological Euler insulators and proposed a theoretical framework
to achieve an all-round characterization of their topological features,
entanglement properties and bulk-boundary correspondence. Through
a series of explicit model studies, we revealed non-Hermiticity induced
phase transitions between topologically distinct Euler insulators,
which are mediated by the formation of gapless intermediate phases.
Non-Hermitian effects were further found to sustain topological insulators
with large Euler class and induce anomalous edge-band overlaps in
both the energy and entanglement spectra. Our findings not only extend
the study of Euler band topology to non-Hermitian systems, but also
unveil unique aspects of topological Euler insulators beyond the Hermitian
limit. 

In future work, it would be interesting to explore the Euler topology
in non-Hermitian systems with more than three bands or fully gapless
spectra \cite{Pan2026PRB}, in higher spatial dimensions \cite{TEI27},
and under external drivings \cite{Pan2025CPL}. In experiments, Hermitian
Euler insulators have been considered in a broad range of quantum
and classical setups, including superconducting qubits \cite{TEI04},
optical lattices \cite{TEI06}, electrical circuits \cite{TEI08},
trapped ions \cite{TEI10}, and acoustic metamaterials \cite{TEI22}.
In many of these systems, non-Hermitian effects like gain and loss
can be flexibly engineered. Therefore, we expect the models we proposed
and the observations of their non-Hermitian topological Euler phases
to be within reach in near-term experiments.

\begin{acknowledgments}
	This work is supported by the National Natural Science Foundation of China (Grant No.~12275260), the Shandong Provincial	Natural Science Foundation (Grant No.~ZR2026QB21), and the Fundamental Research Funds for the Central Universities (Grant No.~202364008).
\end{acknowledgments}

\end{document}